\documentclass[aps, pra,reprint, showpacs,nofootinbib,
superscriptaddress,10pt,floatfix,longbibliography]{revtex4-2}
\usepackage[utf8]{inputenc}
\usepackage{amsmath, amssymb, braket, mathtools, amsthm, MnSymbol,marvosym,xfrac}
\usepackage{graphicx}
\usepackage{ragged2e}
\usepackage[ruled,vlined]{algorithm2e}
\usepackage{bm}
\usepackage[normalem]{ulem}
\usepackage{physics}
\usepackage{hyperref}
\usepackage{orcidlink}
\usepackage[caption=false]{subfig} 
\usepackage{xcolor}     
\usepackage{microtype}      
\usepackage{natbib}
\usepackage{ulem}
\usepackage{soul}
\usepackage{etoolbox}    
\AtBeginEnvironment{thebibliography}{%
  \sloppy
  \setlength\emergencystretch{1em}%
}
\AtEndEnvironment{thebibliography}{\fussy}

\allowdisplaybreaks
\SetKw{KwStep}{step}

\newcommand{\grp}{\gamma_{\scalebox{0.5}{$\rm RP$}}}
\newcommand{\hrp}{H_{\scalebox{0.5}{$\rm RP$}}(G)} 
\newcommand{\lrp}{\widetilde{L}_{\scalebox{0.5}{$\rm RP$}}(G)} 
\newcommand{\pcsadd}{Center for Theoretical Physics of Complex Systems, Institute for Basic Science(IBS), Daejeon 34126, Republic of Korea}

\begin{document}

\title{Marked vertex search on disordered graphs with Rosenzweig–Porter phases}

\author{Sabyasachi Chakraborty\orcidlink{0009-0001-8084-0565}}
\email{sabyasachi.sch@gmail.com}
\affiliation{Department of Physics, Indian Institute of Technology Kharagpur, Kharagpur, West Bengal 721302, India}

\author{Tilen \v{C}ade\v{z}\orcidlink{0000-0002-5343-4086}}
\email{hamilton.ihdt@gmail.com}
\affiliation{Asia Pacific Center for Theoretical Physics, Pohang, Gyeongbuk 37673, Republic of Korea}
\affiliation{\pcsadd}

\author{Sonjoy Majumder\orcidlink{0000-0001-9131-4520}}
\email{sonjoym@phy.iitkgp.ac.in}
\affiliation{Department of Physics, Indian Institute of Technology Kharagpur, Kharagpur, West Bengal 721302, India}

\author{Rohit Kishan Ray\orcidlink{0000-0002-5443-4782}}
\email{rkray@vt.edu}
\affiliation{Department of Materials Science and Engineering, Virginia Tech, Blacksburg, VA 24061, USA}
\affiliation{\pcsadd}

\begin{abstract} 
Quantum marked vertex search algorithms are known to outperform their classical counterparts, yet their behavior in the presence of disorder remains largely unexplored. Here, we address this gap by studying marked vertex search on disordered random graphs. To introduce disorder, we implement the Rosenzweig–Porter (RP) model, a random matrix ensemble with tunable ergodic, non-ergodic extended, and localized phases, on Erd\H{o}s--R\'enyi (ER) graphs. This produces a doubly random system where ER graph connectivity randomizes which interactions exist, while RP disorder controls their strength and `on-site' potentials, providing a two-parameter framework to study quantum dynamics on disordered networks. First, we show that the characteristic Wigner–Dyson-to-Poisson spectral crossover of the RP ensemble survives under graph constraints across the sparse-to-dense range, and we derive an analytical estimate for the finite-size localization boundary that shifts systematically with the graph edge probability $p$, consistent with a resonant-hybridization argument. Thereafter, using this disordered graph ensemble, we study the marked vertex search problem and find that search performance tracks the underlying quantum phase directly. Counterintuitively, the ergodic phase, despite supporting fast transport, yields lower success probability than the localized phase, which achieves high success probability at the cost of significantly longer search times. These results establish a direct and quantitative link between random matrix disorder on graphs and the performance of continuous-time quantum walk search, and suggest that disorder, rather than being merely an obstacle, can be exploited as a tunable parameter in quantum search protocols.

\end{abstract}

\maketitle

\section{Introduction}

Disorder influences the dynamics of quantum systems given its significant impact on transport, localization, and quantum information processing. Continuous-time quantum walks (CTQWs)~\cite{farhi1998quantum, Kempe01072003, Portugal2013} provide a natural framework for investigating this interplay, since changes in the spectral and eigenstate properties of the underlying graph Hamiltonian ($H$) are directly reflected in the walker dynamics. As a quantum analog of classical random walks, CTQWs replace classical probability distributions with complex probability amplitudes, and diffusive stochastic dynamics is replaced by unitary evolution on a graph. The walker evolves in the graph according to the Schr\"odinger equation, $i\frac{d}{dt}\ket{\psi(t)} = H\ket{\psi(t)}$, where $H$ encodes the connectivity of the underlying network~\cite{farhi1998quantum}. Within this framework, CTQWs provide a natural setting for spatial quantum search, a graph problem in which the objective is to identify a marked vertex through coherent evolution on the underlying network~\cite{ChildsPRA_2004, sChakraborty_prl2016, chakraborty_optimality_PRA}. 

In their seminal work, \citet{ChildsPRA_2004} showed that CTQW-based spatial search can locate a marked vertex among $N$ graph vertices in $\mathcal{O}(\sqrt{N})$ time for suitable graphs, such as complete graphs, hypercubes, and $d$-dimensional lattices with $d>4$, giving a quadratic speedup over the classical $\mathcal{O}(N)$ search. Subsequent studies~\cite{sChakraborty_prl2016, chakraborty_optimality_PRA, ApersPRL_2022} have shown that the efficiency of this search process is strongly governed by the spectral structure of the graph Hamiltonian, including eigenvalue gaps and eigenstate overlaps with the marked vertex. Since the underlying graph topology directly shapes these spectral properties, features such as connectivity, sparsity, disorder, clustering, and spectral heterogeneity can significantly modify transport and search performance. Highly connected or ergodic graphs generally support rapid amplitude spreading, whereas sparse or disordered graphs may induce localization effects that strongly alter the search dynamics~\cite{MULKEN_PRE2007, MULKEN201137, sChakraborty_prl2016}. Understanding the interplay between graph structure, spectral properties, coherent transport, and localization is therefore essential for characterizing CTQW-based quantum search on complex networks~\cite{ChildsPRA_2004, chakraborty_optimality_PRA, ApersPRL_2022}.

In solid-state quantum systems and complex networks, imperfections naturally arise through random onsite energies, fluctuating coupling strengths, missing links, or uncontrolled environmental variations~\cite{anderson1958absence, LeeRevModPhys, Razzolie23010085, roth2023tuning, kurt2023quantum, longhi2023anderson, khodadad2025impact}. Such disorders can reshape the spectral structure of the Hamiltonian and modify the propagation of quantum amplitudes. A well-known example is Anderson  localization~\cite{anderson1958absence, DJThouless_1970}, where in three or more spatial dimensions, sufficiently strong disorder suppresses wave spreading through destructive interference and produces spatially localized eigenstates; in one and two dimensions, arbitrarily weak disorder eventually localizes all single-particle states~\cite{abrahams_1979_scaling}. This scenario becomes significantly more complex in graph-based systems because the network topology determines the hopping directions~\cite{Chen_PRL2026, SahaRoy_PRB2026, Lucas_PRE2025, Sierant_SciPostPhys2023, Tikhonov_PRB2016}. Depending on the interplay between connectivity and disorder, eigenstates can be fully extended and ergodic, localized around a small subset of vertices, or show intermediate non-ergodic extended behavior~\cite{Luca_PRL2014, Pino_PRR2020, Tilen_PRB2023, Cugliandolo_PRB2024, SahaRoy_PRB2026}.

CTQW-based spatial search has seen extensive progress~\cite{ChildsPRA_2004, sChakraborty_prl2016, chakraborty_optimality_PRA}, yet it remains largely unexamined in the presence of the disorder-induced localization~\cite{anderson1958absence, DJThouless_1970} and random-matrix phenomena~\cite{Luca_PRL2014, Pino_PRR2020, Cugliandolo_PRB2024, SahaRoy_PRB2026} that are known to reshape the underlying graph spectra. Existing CTQW search studies primarily neglect onsite disorder and relate the search dynamics to the spectral properties of the underlying graph~\cite{sChakraborty_prl2016, chakraborty_optimality_PRA}. Even in studies involving random graphs, randomness usually enters through the graph topology, while the hopping amplitudes remain deterministic once the graph is specified. In contrast, most of the studies of disorder in quantum walks~\cite{Yin_PRA2008, Benedetti_PRA2016, chawla2019quantum} have mainly focused on transport, spreading, return probability, and Anderson-type localization, rather than the algorithmic problem of finding a marked vertex. Although a few related studies have explored noisy CTQW spatial search and search on nontrivial graph topologies~\cite{Cattaneo_PRA2018, Malmi_PRR2022}, a systematic understanding of marked-vertex search in the presence of Hamiltonian disorder, eigenstate localization, and random graph topology is still lacking. In particular, CTQW search is known to be highly sensitive to spectral gaps and eigenstate overlaps with the marked vertex, so it is not clear whether efficient transport necessarily leads to better search, or whether localization can favor amplitude accumulation near the marked vertex. This tension defines the central question of the present work: does disorder degrade CTQW-based search, or can it instead be harnessed---at some cost---as a resource that reshapes search performance? 

A starting point for investigating disorder is to introduce it on a simplistic model such as a line graph and study the resulting marked-vertex search dynamics. However, one-dimensional graphs generally do not support optimal spatial search~\cite{ChildsPRA_2004} and provide only limited spectral and eigenstate structure for capturing the broader effects of disorder. To study disorder in a controlled way, one therefore needs a model in which the structure of the quantum states can be tuned systematically from ergodic to localized behavior. The Rosenzweig--Porter (RP)~\cite{rosenzweig_1960_repulsion,kravtsov_2015_random, Facoetti2016, Truong2016, Monthus2017, vonSoosten2019, Pino2019, DeTomasi2019, Berkovits2020,
khaymovich_2021_dynamical, Zhang2023, Buijsman_PRB2024, cadez_2024_rosenzweig, Zhang2026} ensemble provides such a random-matrix framework. In the RP model, random onsite potentials compete with disorder-scaled off-diagonal couplings, while the parameter $\grp$ controls the hybridization between basis states. 
Varying $\grp$ drives the system through ergodic, non-ergodic extended, and localized regimes, thereby connecting spectral statistics, eigenstate structure, and localization within a single tunable framework. To use RP physics as a control parameter for CTQW-based marked-vertex search on networks, the RP disorder structure must be embedded in a graph-constrained model, where hopping is restricted to the edges of the underlying graph.

This motivates the construction used in the present work, where we implement the RP ensemble on Erd\H{o}s--R\'enyi (ER)~\cite{erdos1959random, erdos1959evolution} random graphs by retaining random diagonal onsite terms while constraining the off-diagonal RP couplings to the edges of the ER graph. The ER graph selects which hopping channels are present, while the RP scaling controls the strength of the allowed random hoppings, and the diagonal terms provide onsite disorder. The resulting construction therefore defines a graph-constrained RP ensemble, where network sparsity and disorder strength can be tuned independently. This doubly random system provides a natural testbed for studying how random connectivity and spectral correlations jointly reshape localization, and, in turn, CTQW-based marked-vertex search. 

Our analysis shows that the graph-constrained RP ensemble preserves the characteristic crossover~\cite{rosenzweig_1960_repulsion,kravtsov_2015_random,
khaymovich_2021_dynamical,Buijsman_PRB2024, cadez_2024_rosenzweig} from Wigner--Dyson to Poisson level statistics as the parameter $\grp$ is increased. Using a resonant-hybridization argument, we further obtain a finite-size localization boundary that depends explicitly on both the ER edge probability $p$ and the system size $N$. This boundary recovers the conventional RP threshold in both the fully connected and thermodynamic limits. We then investigate CTQW-based marked-vertex search and identify a clear trade-off between success probability and runtime. The ergodic regime supports faster search dynamics but a lower peak success probability, whereas increasing localization enhances the marked-state amplitude at the cost of longer search times. These findings show that controlled disorder can act as a tunable resource in CTQW search, balancing transport efficiency against marked-state amplitude concentration.

The rest of the paper is organized as follows. In Section~\ref{sec:RP_model}, we review the RP model and summarize the characteristic phases. In Section~\ref{sec:RP_on_graphs}, we introduce the graph-constrained RP ensemble on Erd\H{o}s--R\'enyi random graphs, where the edge probability $p$ controls the network sparsity and $\grp$ controls the strength of the allowed random couplings. We then validate this construction using adjacent-gap-ratio statistics and derive an analytical estimate for the $p$-dependent finite-size localization crossover. In Section~\ref{sec:search_RP_graph}, we study marked-vertex search by continuous-time quantum walks in the presence of disorder. We first recall the standard spatial-search formulation in Section~\ref{sec:search_stnd_graph} and the \citet{chakraborty_optimality_PRA} spatial search framework in Section~\ref{sec:CNR_algo}. We then analyze a disordered line graph in Section~\ref{sec:disord_linegraph} for understanding how local disorder modifies the spectral structure relevant for search. At last, in Section~\ref{sec:search_tradeoff}, we investigate marked-vertex search in the full RP--ER ensemble and demonstrate the trade-off between success probability and runtime. Finally in Section~\ref{sec:conclusion}, we summarizes our findings and outlines future perspectives.

\section{The Rosenzweig--Porter model and its phases}
\label{sec:RP_model}

The RP ensemble provides a random-matrix framework in which the competition between diagonal disorder and random off-diagonal coupling can be tuned continuously~\cite{rosenzweig_1960_repulsion,kravtsov_2015_random,
khaymovich_2021_dynamical,Buijsman_PRB2024, cadez_2024_rosenzweig}.
For the real-symmetric version considered in this work, the RP Hamiltonian is written as
\begin{equation}
    H_{\mathrm{RP}} = H_{0} + \frac{\nu}{N^{\grp/2}}\,M ,
\label{eq:RP_Hamiltonian}
\end{equation}
where $N$ is the matrix dimension and $\grp\geq 0$ is the RP tuning parameter. The matrix $H_{0}$ is diagonal,
\begin{equation}
    H_{0}=\mathrm{diag} \left(\epsilon_{1},\epsilon_{2},\ldots,\epsilon_{N}\right),
\end{equation}
with the onsite energies $\epsilon_i$ sampled independently from a Gaussian distribution with zero mean and unit variance. The matrix $M$ is drawn from the Gaussian orthogonal ensemble (GOE) while the parameter $\nu=\mathcal{O}(1)$ sets the overall strength of the random off-diagonal couplings. Since an overall variance factor can be absorbed into $\nu$, different normalization conventions for the GOE do not alter the phase structure of the model.

A convenient numerical construction of $M$ is obtained from a real random matrix $A$ as
\begin{equation}
    M=\frac{1}{2}\left(A+A^{T}\right),
    \label{eq:GOE_construction}
\end{equation}
where the entries of $A$ are independent Gaussian random variables with zero mean and variance $\frac{1}{2}$. Eq.~\eqref{eq:GOE_construction} guarantees that $M$ is real and symmetric, we further impose the condition $M_{ii}=0$ so that the diagonal disorder is contained entirely in $H_0$. Also, one may generate the upper-triangular elements independently and impose $M_{ji}=M_{ij}$. 

The factor $N^{-\grp/2}$ in Eq.~\eqref{eq:RP_Hamiltonian} controls the strength of the random-matrix contribution relative to the diagonal disorder as the system size increases. This competition gives rise to three asymptotic regimes in the conventional RP ensemble~\cite{kravtsov_2015_random, khaymovich_2021_dynamical, Buijsman_PRB2024, cadez_2024_rosenzweig}.

For $0\leq \grp<1$, the RP model is in the \emph{ergodic extended phase}, where eigenstates are fully hybridized and the energy levels follow Wigner--Dyson statistics~\cite{Altland_PRE_1997, Serbyn_PRB2016, Tekur_PRB2018, Buijsman_PRL2019, Tian_PRL2024, Buijsman_PRB2024}. In the interval $1<\grp<2$, the system enters a \emph{nonergodic extended}, or fractal, phase~\cite{kravtsov_2015_random, Bogomolny_PRE2018}. In this regime, the eigenstates are extended but nonergodic, and the short-range spectral correlations remain asymptotically Wigner--Dyson. Level-spacing statistics alone do not generally distinguish ergodic from nonergodic extended phases in the thermodynamic limit~\cite{kravtsov_2015_random, Bogomolny_PRE2018, Buijsman_PRB2024}. For $\grp>2$, the system is in the \emph{localized phase}, characterized by asymptotically uncorrelated energy levels and Poisson statistics~\cite{Altland_PRE_1997, Serbyn_PRB2016, Tekur_PRB2018, Buijsman_PRL2019, Tian_PRL2024, Buijsman_PRB2024}. The points $\grp=1$ and $\grp=2$ mark the ergodic-to-nonergodic-extended and extended-to-localized transitions, respectively, with $\grp=2$ also marking the change from Wigner--Dyson to Poisson level statistics~\cite{Altland_PRE_1997, kravtsov_2015_random, Bogomolny_PRE2018}.

To characterize the change in spectral correlations, we use the ratio of consecutive level spacings~\cite{Oganesyan2007, atas2013distribution}, which does not require spectral unfolding. Let $\left\{E_n\right\}_{n=1}^{N}$ denote the ordered eigenvalue set, with $E_1\leq E_2\leq\cdots\leq E_N$. The nearest-neighbor level spacings are defined as $s_n=E_{n+1}-E_n$ for $n=1,\ldots, N-1$. The ratio of two consecutive spacings is then defined by
\begin{equation}
    r_n = \frac{\min(s_n,s_{n+1})} {\max(s_n,s_{n+1})},
    \qquad n=1,\ldots,N-2.
\label{eq:spacing_ratio}
\end{equation}

The disorder-averaged spacing ratio is defined as $\expval{r} = \expval{r_n}_{n,\mathrm{dis}}$, where the average is taken over the full spectrum and over independent realizations of the disorder. For the GOE universality class, Wigner--Dyson statistics give $\expval{r}\simeq 0.5307$ (GOE limit), whereas uncorrelated Poisson statistics produces $\expval{r}\simeq 0.3863$ (Poisson limit)~\cite{Oganesyan2007, atas2013distribution}. For finite systems, intermediate values generally represent a crossover between these two limiting behaviors.

\section{Rosenzweig--Porter on graphs}
\label{sec:RP_on_graphs}
The conventional RP ensemble introduced in the previous section corresponds to an effectively fully connected graph. For an undirected graph $G=(V, E)$, the vertex set is $V={v_1,v_2,\ldots,v_N}$, and the edge set $E=\{e_{i,j} \textbf{ s.t. } e_{i,j} \text{ connects } (v_i, v_j)\}$ is the set of undirected edges. In a fully connected graph, an edge exists between every pair of distinct vertices.

In the present work, to introduce a tunable graph constraint, we will use the ER random graph $G(N,p)$~\cite{erdos1959random, erdos1959evolution}, where $N$ is the number of vertices, and $p$ is the probability that there is an edge between any two of them. Incorporating the connectivity constraints of the ER graph on top of the conventional RP ensemble will give us a unique doubly random system. In the next subsection, we will introduce the model in details.

\subsection{The Model}
For an ER random graph $G(N,p)$ with vertex set $V=\{v_1,v_2,\ldots,v_N\}$, the corresponding adjacency matrix is
\begin{equation}
A_{ij} =
\begin{cases}
1, & \text{with probability } p \quad (i\neq j),\\
0, & \text{otherwise}.
\end{cases}
\end{equation}

Since the graph is undirected, $A_{ij}=A_{ji}$.

The degree, $d_i$, of a vertex $v_i$ is the number of edges connected to it, and can be written in terms of the adjacency matrix as $d_i=\sum_{j=1}^{N} A_{ij}$. The corresponding degree matrix is the diagonal matrix $D_{ij}=d_i\delta_{ij}$, where $\delta_{ij}$ is the Kronecker delta. The graph Laplacian is then defined as $L=D-A$~\cite{Ray_PRE2025}.

The ER graph specifies only the connectivity of the system. To introduce RP like disorder, we assign random onsite energies to the vertices and random hopping amplitudes to the edges. The diagonal disorder is described by
\begin{equation}
V_{ij}=\varepsilon_i\delta_{ij},
\label{eq:diag_dis}
\end{equation}
where the onsite potentials are independently drawn from a normal distribution $\varepsilon_i \sim \mathcal{N}\!\left(0,(2N)^{-1}\right)$. 

\begin{figure}[t!]
    \centering
    \includegraphics[width=\linewidth]{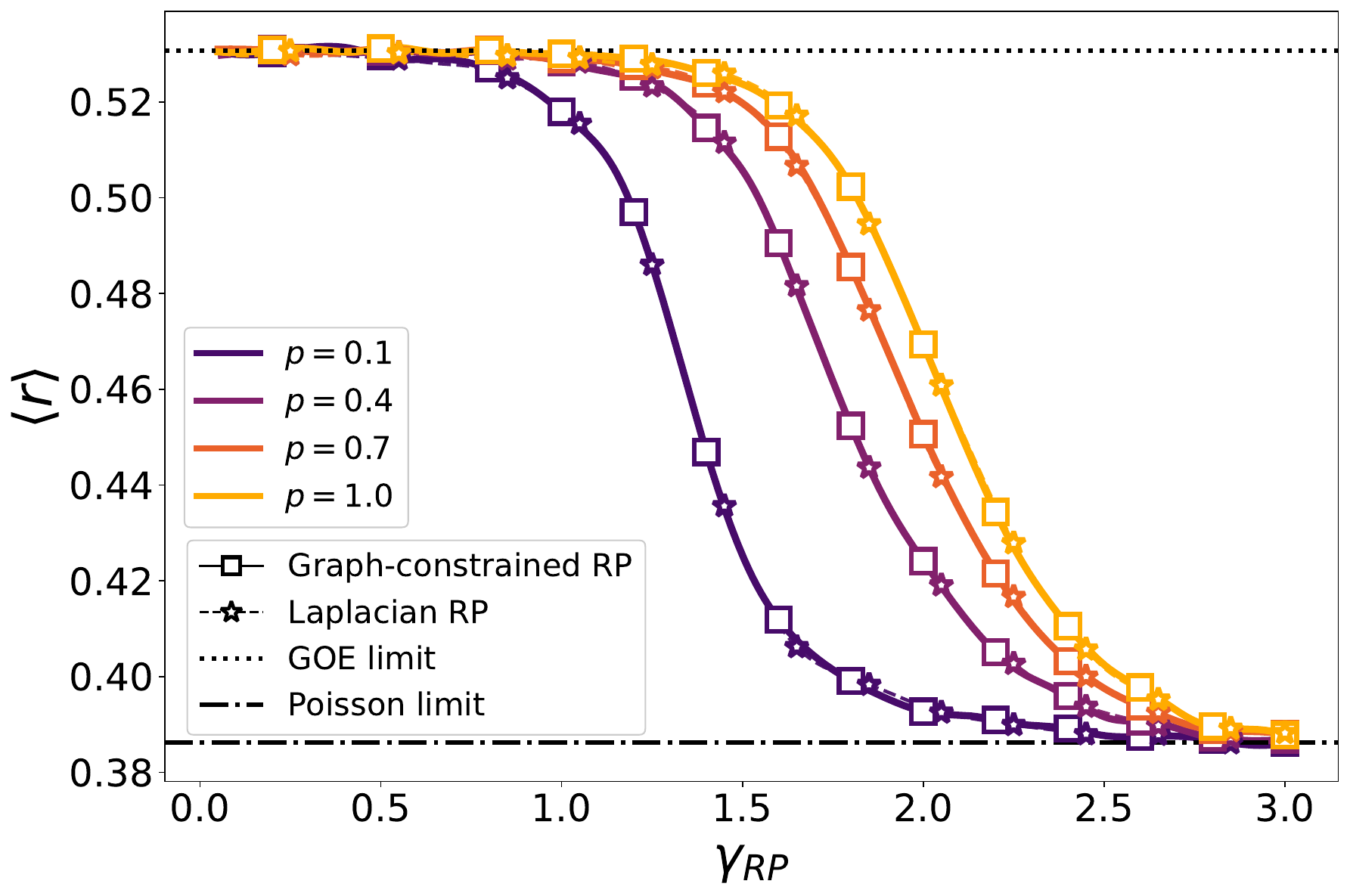}
    \caption{Disorder-averaged spacing ratio $\expval{r}$ as a function of the RP off-diagonal scaling parameter $\grp$ for ER graphs with $N=1000$ and edge probabilities $p=0.1,0.4,0.7,1.0$. Different colors denote different values of $p$. Square markers correspond to the graph-constrained RP Hamiltonian Eq.~\eqref{eq:RP_Hamil}, while star markers correspond to the Laplacian-based RP operator Eq.~\eqref{eq:RP_laplacian}. In both constructions, the ER adjacency matrix acts as a sparsity mask that determines the locations of the nonzero RP-scaled couplings. The horizontal dotted and dash-dotted lines indicate the GOE and Poisson limits, respectively.}
    \label{fig:Rp_lvlspacing}
\end{figure}

This normalization keeps the diagonal contribution comparable to the off-diagonal terms in the large-$N$ limit. The off-diagonal disorder is introduced through random couplings, but only along the edges of the ER graph. For $i\neq j$, the off-diagonal disorder is introduced through random couplings along the edges of the ER graph as:

\begin{equation}
    W = A\circ X, ~~\text{with } X_{ij}\sim
\mathcal{N}\!\left(0,\frac{1}{4N^{\grp+1}}\right)
\label{eq:RP_ER_dis}
\end{equation}
where $\circ$ denotes Hadamard product and $X$ is the symmetric random coupling.

\begin{figure*}[t!]
  \centering
 \subfloat[ \label{fig:RP_phasechange1}$p=0.1$]{\includegraphics[width=0.68\columnwidth]{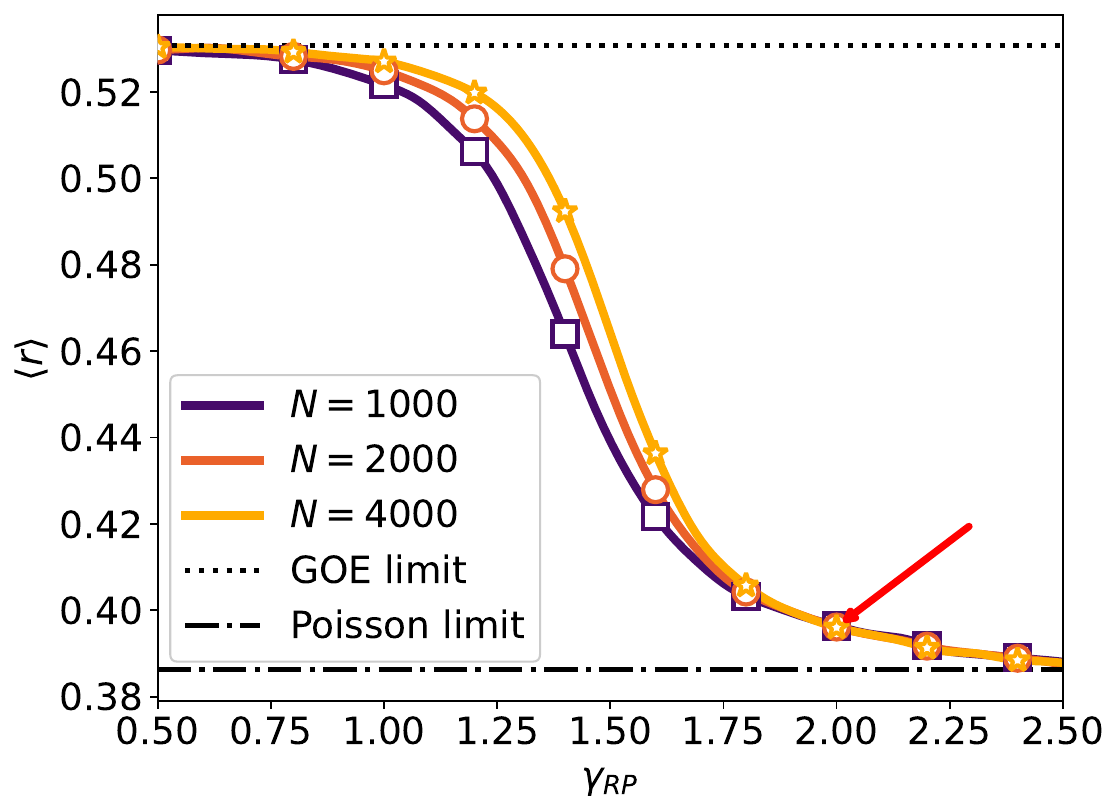}} 
 \subfloat[ \label{fig:RP_phasechange2}$p=0.5$]{\includegraphics[width=0.68\columnwidth]{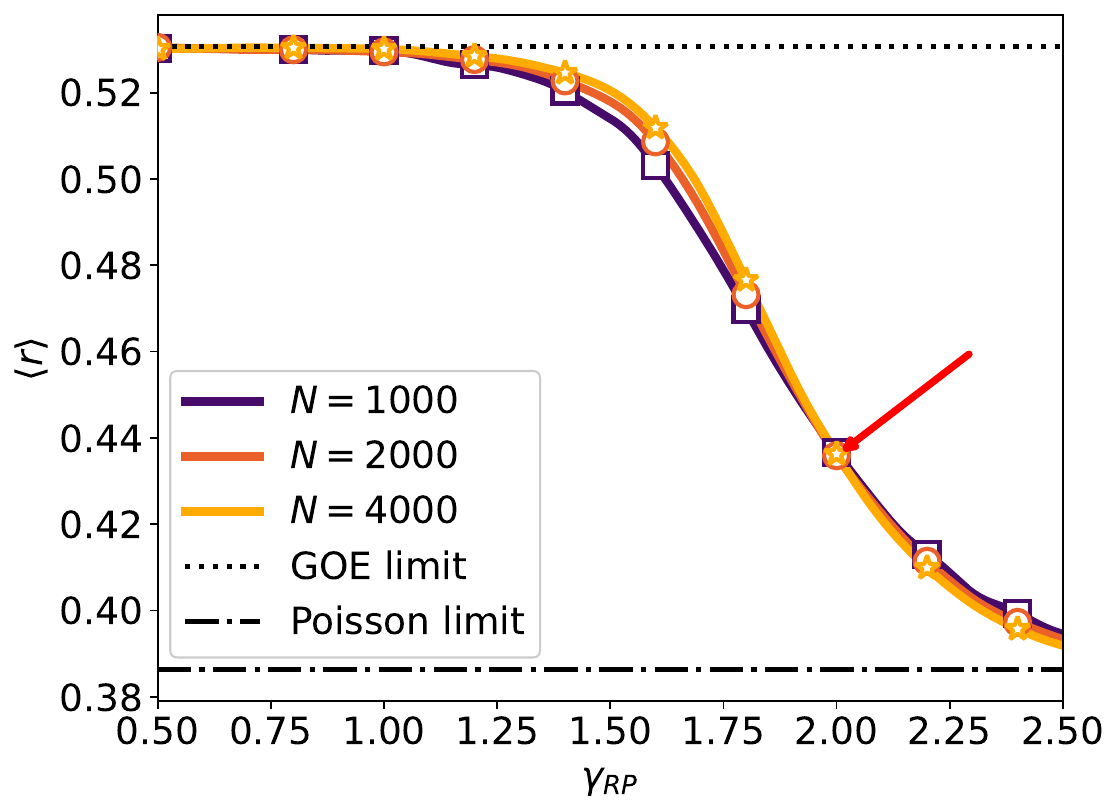}} 
 \subfloat[ \label{fig:RP_phasechange3}$p=1.0$]{\includegraphics[width=0.68\columnwidth]{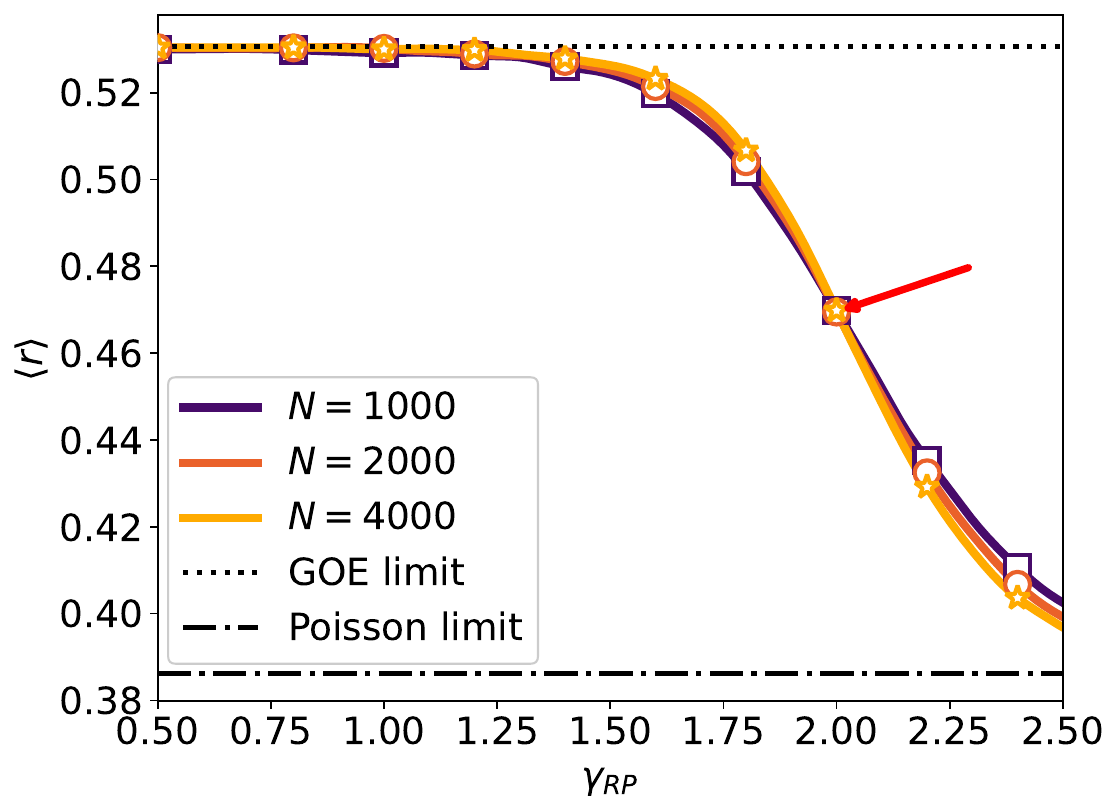}}
  \caption{Finite-size behavior of the disorder-averaged spacing ratio $\expval{r}$ as a function of the RP off-diagonal scaling parameter $\grp$ for the graph-constrained RP Hamiltonian at representative edge probabilities (a) $p=0.1$, (b) $p=0.5$, and (c) $p=1.0$. Results are shown for $N=1000,2000,$ and $4000$. The curves display a crossover from GOE-like statistics at small $\grp$ to Poisson-like statistics at large $\grp$. The horizontal dotted and dash-dotted lines denote the GOE and Poisson limits, respectively. The crossover point is used to estimate the critical scaling parameter $\gamma_c$.}
  \label{fig:RP_phasechange}
\end{figure*}

Thus, the adjacency matrix acts as a mask for the RP ensemble. If two vertices are not connected in the ER graph, then $A_{ij}=0$ and the corresponding hopping matrix element is exactly zero. If an edge is present, $A_{ij}=1$, the hopping amplitude is drawn from a Gaussian distribution whose variance is controlled by the RP exponent $\grp$. By construction, $W_{ii}=0$ and $W_{ij}=W_{ji}$, so that the hopping matrix is real and symmetric. In this way, the two parameters $p$ and $\grp$ control two different aspects of the model; $p$ fixes the sparsity of the underlying graph, while $\grp$ fixes the strength of the allowed random hoppings.

For a given graph $G$, we define the graph-constrained RP Hamiltonian $\hrp$, whose off-diagonal structure follows the topology of $G$. The corresponding graph-constrained RP Laplacian is denoted by $\lrp$. The Hamiltonian $\hrp$ is defined as,

\begin{equation}
\hrp=V+W,
\label{eq:RP_Hamil}
\end{equation}

or, equivalently,
\begin{equation}
\left[\hrp\right]_{ij} = \varepsilon_i\delta_{ij} + A_{ij}X_{ij}.
\label{eq:RP_Hamil1}
\end{equation}

This Hamiltonian combines onsite disorder with RP-scaled random hopping constrained by the underlying graph $G$. The ER graph determines which off-diagonal matrix elements are present, while the RP scaling controls their magnitude. In the fully connected case, $p=1$, all off-diagonal couplings are allowed, and the model reduces to the conventional Rosenzweig--Porter ensemble.

For constructing a graph Hamiltonian, we consider same protocol used for Laplacian. Starting from the weighted adjacency matrix $W$, we define the corresponding weighted degree matrix $D_{ij} = \delta_{ij}\sum_{k=1}^{N}W_{ik}$.

The Laplacian-type RP operator is then written as
\begin{equation}
\lrp=D-W+V.
\label{eq:RP_laplacian}
\end{equation}
The term $D - W$ is the Laplacian of the weighted graph associated with the weight of the RP-scaled edge, and the matrix $V$ adds the diagonal disorder on site. Therefore, compared to Hamiltonian $\hrp$, the operator $\lrp$ contains an additional diagonal contribution determined by the weighted connectivity of each vertex.

The two operators $\hrp$ and $\lrp$ represent two distinct ways of embedding RP disorder into a random graph. $\hrp$ is a graph-constrained Hamiltonian, in which the off-diagonal RP couplings are restricted by the connectivity of the underlying graph, while $\lrp$ is the corresponding weighted graph Laplacian. Although they differ structurally, because the same RP weights are embedded on the same underlying graph, we find that they produce nearly identical spectral statistics across the localization crossover.

\subsection{Spectral validation of the model}

To characterize the spectral properties of the graph-constrained RP model, we use the disorder-averaged spacing ratio (see Eq.~\eqref{eq:spacing_ratio}). FIG~\ref{fig:Rp_lvlspacing} shows $\expval{r}$ as a function of the RP scaling parameter $\grp$ for ER graphs with $N=1000$ and several edge probabilities $p$. The results are obtained by averaging over $1500$ independent graphs and disorder realizations. For small $\grp$, the off-diagonal random couplings are sufficiently strong to hybridize the eigenstates, and the level statistics remain close to the GOE value~\cite{kravtsov_2015_random, Bogomolny_PRE2018, Buijsman_PRB2024, Tilen_PRB2023}. As $\grp$ increases, the RP-scaled hoppings become weaker, and the system crosses over towards Poisson statistics, indicating localization~\cite{kravtsov_2015_random, Bogomolny_PRE2018, Buijsman_PRB2024, Tilen_PRB2023}. The onset of the crossover from GOE-like to Poisson-like statistics depends strongly on the sparsity of the underlying graph. For low $p$ values, deviations from the GOE-like regime appear at lower $\grp$; for example, the crossover begins around $\grp \approx 1.0$ for $p=0.1$, whereas for $p=1$ it starts near $\grp \approx 1.5$. As $p$ increases, the crossover also becomes sharper. However, as we will see below, this transition away from GOE statistics is not the same as the universal crossover behavior seen in RP literature. Rather, it reflects a finite-size connectivity effect, whereby the reduced number of available hopping channels suppresses eigenstate hybridization, causing `less' availability of ergodic regime. FIG~\ref{fig:Rp_lvlspacing} also compares $\hrp$ (see Eq.~\eqref{eq:RP_Hamil}) with the $\lrp$ (see Eq.~\eqref{eq:RP_laplacian}). The two constructions give almost indistinguishable values of $\expval{r}$ over the entire range of $\grp$ and for all values of $p$ considered. This provides a numerical consistency check of the model.

We further examine the finite-size behavior in FIG.~\ref{fig:RP_phasechange}, where $\expval{r}$ is plotted as a function of $\grp$ for $N=1000,2000,$ and $4000$ at the representative edge probabilities $p=0.1,0.5,$ and $1.0$. Data are averaged on $1500$ independent realizations for $N=1000$, on $1000$ realizations for $N=2000$, and on $500$ realizations for $N=4000$. For each value of $p$, the curves display a clear crossover between the GOE and Poisson limits. We estimate the corresponding finite-size crossover $\gamma_c$ as the value of $\grp$ where the spacing-ratio curves for $N=1000,2000,$ and $4000$ are closest to one other, namely where their mutual separation is minimized at fixed edge probability $p$. The extracted critical value of $\gamma_c$ for a wider range of edge probabilities, $p=0.1$ to $1.0$, are shown in FIG.~\ref{fig:tuned_gamc_all} and are consistent with the transition at $\gamma = 2$.

These numerical results validate our construction of the ER-constrained RP model. The parameter $p$ controls the number of available hopping channels, while $\grp$ controls the strength of the allowed hoppings. Their combined effect produces a tunable crossover from an ergodic, GOE-like regime to a localized, Poisson-like regime.

\subsection{Analytical estimate of the $p$-dependent localization threshold}
\begin{figure}[t!]
    \centering
    \includegraphics[width=\linewidth]{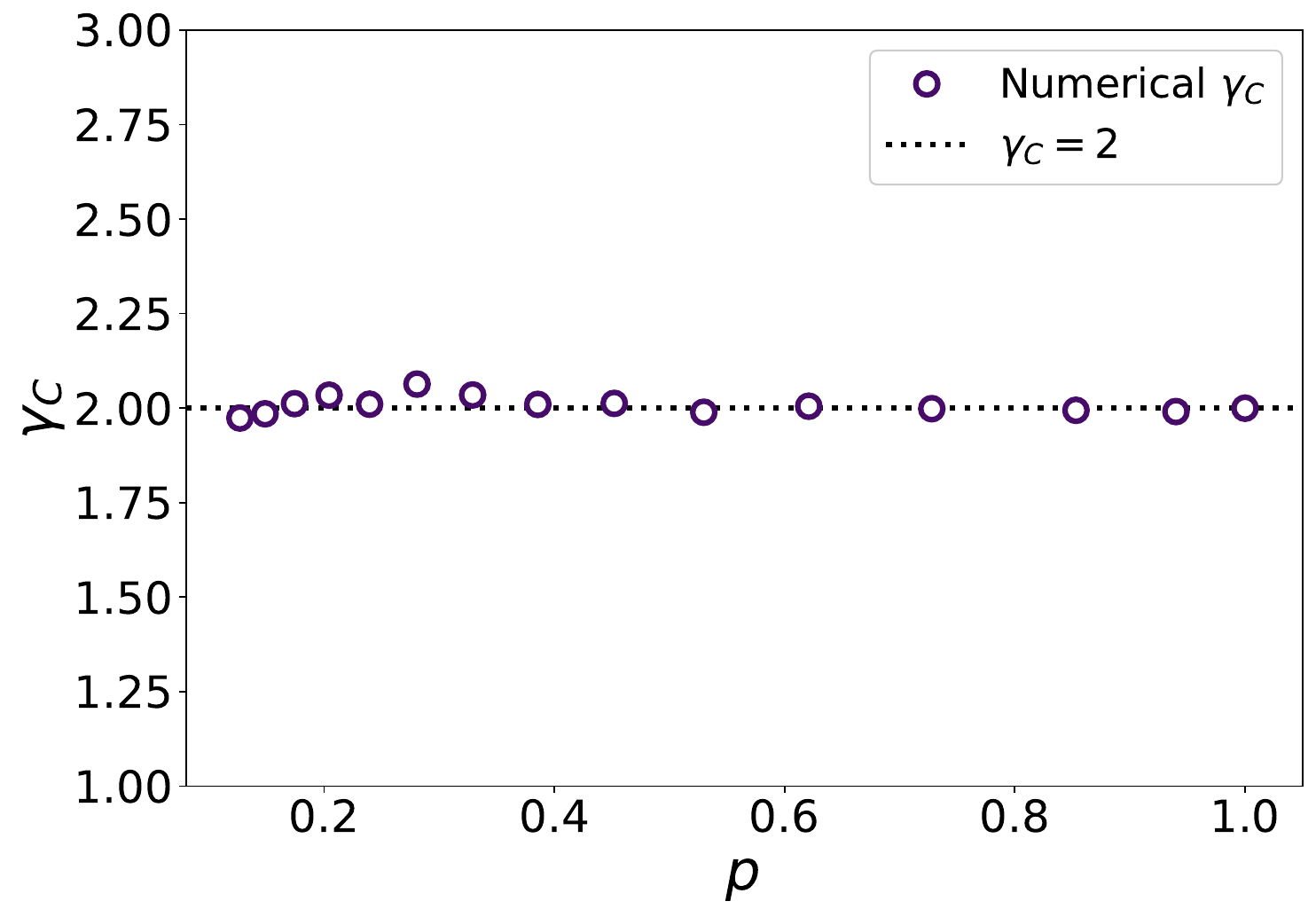}
    \caption{Finite-size crossover $\gamma_c$ as a function of the ER edge probability $p$. For each value of $p$, $\gamma_c$ is identified as the value of $\grp$ at which the adjacent-gap-ratio curves for $N=1000$, $2000$, and $4000$ have the minimum mutual separation. The dotted horizontal line marks the conventional RP localization threshold, $\gamma_c=2$.}
    \label{fig:tuned_gamc_all}
\end{figure}

The numerical results in FIGs.~\ref{fig:RP_phasechange} and~\ref{fig:tuned_gamc_all} show that the localization crossover is controlled by both $\grp$ and $p$. To understand this dependence, we derive the analytical condition under which the off-diagonal couplings scaled by RP cease to efficiently hybridize the nearby basis states. The physical criterion is that, localization occurs when the number of basis states resonantly coupled to a given site no longer grows with the system size, but remains finite in the thermodynamic limit~\cite{Biroli_PRB2021}. In this regime, resonant mixing is suppressed, and the level statistics transitions from GOE-like to Poisson-like behavior~\cite{kravtsov_2015_random, khaymovich_2021_dynamical}.

From the definition of the ER-constrained RP ensemble, a nonzero off-diagonal matrix element has variance $(4N^{\grp+1})^{-1}$ (see Eq.~\eqref{eq:RP_ER_dis}). Therefore, its typical magnitude scales as $|X_{ij}|\sim N^{-(\grp+1)/2}$. On the other hand, the onsite disorder has a characteristic scale $W_{\epsilon}\sim N^{-1/2}$ (see Eq.~\eqref{eq:diag_dis}). A connected pair of sites becomes resonant when the corresponding energy mismatch is smaller than the hopping amplitude, \textit{i.e.},
\begin{equation}
|\epsilon_i-\epsilon_j|\lesssim |X_{ij}|.
\end{equation}

For a connected pair, the estimated probability to satisfy this condition is the corresponding probability to satisfy resonance is:
\begin{equation}
P_{\rm res} \sim \frac{|X_{ij}|}{W_{\epsilon}}
\sim \frac{N^{-\sfrac{(\grp+1)}{2}}}{N^{-1/2}} = N^{-\grp/2}.
\end{equation}
For an ER graph, the average degree is $k\simeq p(N-1)\approx pN$. Thus, the typical number of resonant neighbors connected to a given site is
\begin{equation}
k_{\rm res} \sim k P_{\rm res} 
\sim pN\,N^{-\grp/2}
= p\,N^{1-\sfrac{\grp}{2}}.
\end{equation}

\begin{figure}[ht!]
  \centering
 \subfloat[ \label{fig:rp_phase_diagram1}$N=1000$]{\includegraphics[width=0.8\columnwidth]{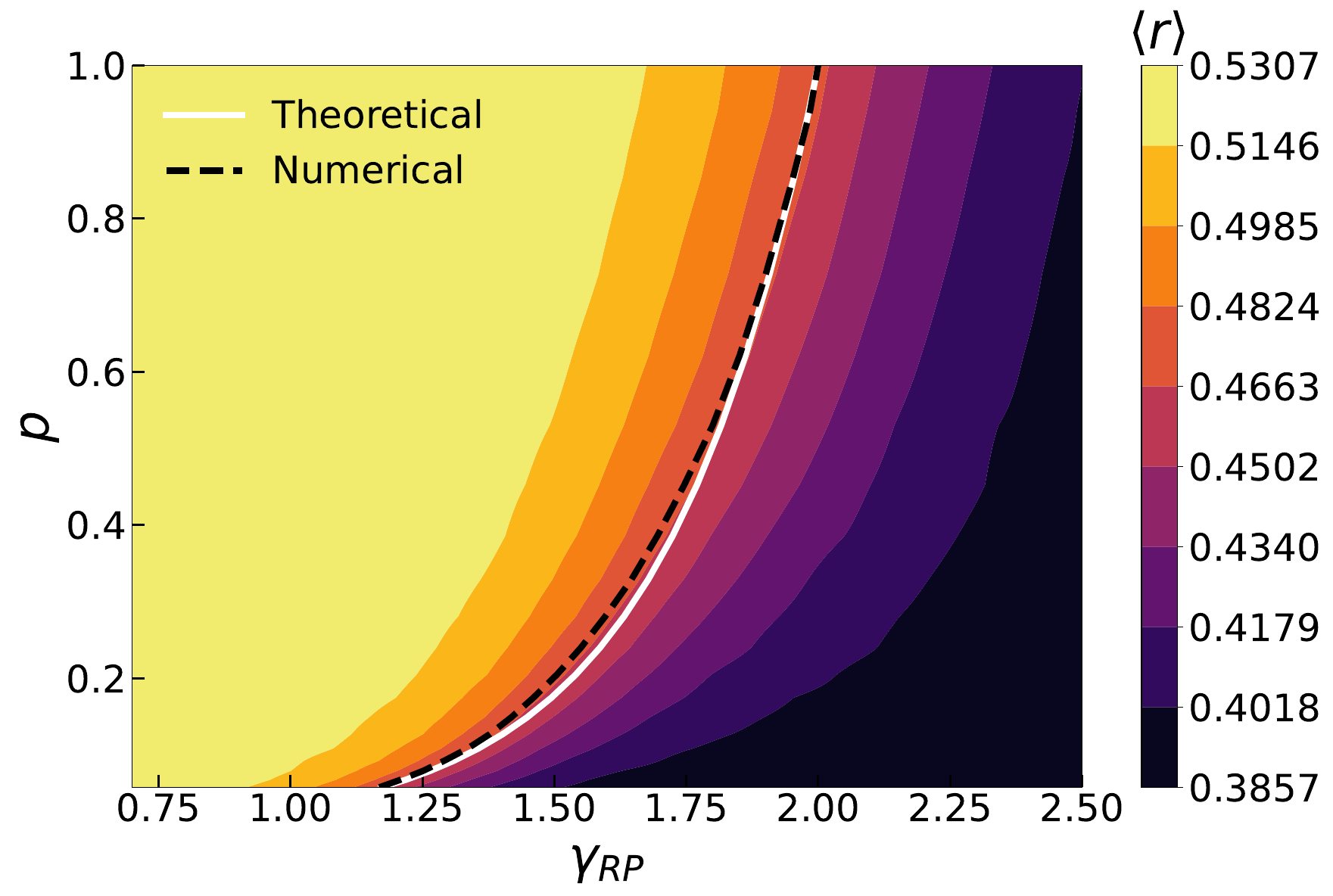}} \\
 \subfloat[ \label{fig:rp_phase_diagram2}$N=4000$]{\includegraphics[width=0.8\columnwidth]{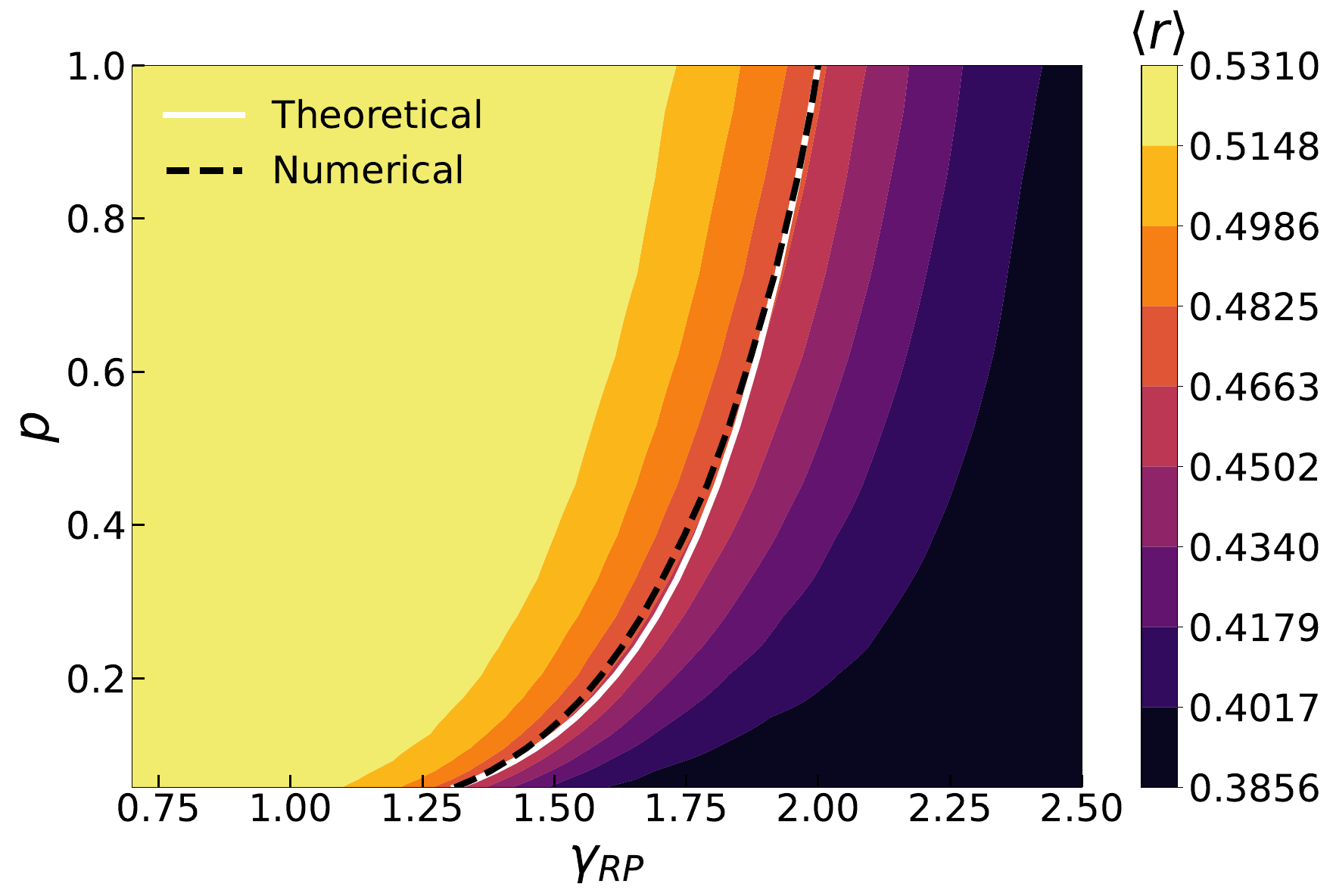}} 
  \caption{Contour plot of the mean adjacent-gap ratio $\expval{r}$ in the $(\grp,p)$ plane for system sizes (a) $N=1000$ and (b) $N=4000$. The color scale denotes the value of $\expval{r}$, with larger values corresponding to GOE-like statistics and smaller values approaching the Poisson limit. The black dashed curve denotes the numerical crossover line obtained from the anchored iso-$\expval{r}$ criterion, $\expval{r(\grp,p)}=\expval{r(\grp=2,p=1)}$. The white curve shows the analytical estimate $\gamma_{\expval{r}}(p,N)=2+2\log p/\log N$, obtained from the effective-resonance condition $k_{\rm res}\sim pN^{1-\grp/2}\sim 1$.}
  \label{fig:rp_phase_diagram}
\end{figure}

Delocalization requires the number of resonantly coupled basis states, $k_{\rm res}$, to grow with the size of the system. However, localization occurs when $k_{\rm res}$ remains finite in the thermodynamic limit~\cite{Biroli_PRB2021}. We therefore estimate the localization limit by imposing the marginal condition $k_{\rm res}\sim \mathcal{O}(1)$, which gives the following.
\begin{equation}
p N^{1-\grp/2}\sim 1 .
\end{equation}
Taking the logarithms of both sides, we obtain the following.
\begin{equation}
\log p+\left(1-\frac{\gamma_{RP}}{2}\right)\log N=0,
\end{equation}
yielding the finite-size effect on $\grp$ at transition
\begin{equation}
\gamma_{\expval{r}}(p,N)=
2+\frac{2\log p}{\log N}.
\label{eq:gamma_p_threshold}
\end{equation}
We use the subscript `$\expval{r}$' here instead of `$\scalebox{0.5}{\rm RP}$' because this expression identifies the transition on the $\expval{r}$ landscape (see Fig.~\ref{fig:rp_phase_diagram}).

Equation~\eqref{eq:gamma_p_threshold} describes the finite-size, $p$-dependent crossover line in the $\expval{r}$ landscape. In the fully connected case, $p=1$, it gives $\gamma_{\expval{r}}(1,N)=2$, consistent with the finite-size crossing value $\gamma_c = 2$ extracted from the minimum mutual separation of the spacing-ratio curves for different system sizes (see Fig.~\ref{fig:tuned_gamc_all}). For $p<1$ and fixed $N$, the reduced number of available hopping channels shifts the crossover line to values $\gamma_{\expval{r}}(p,N)<2$. This shift is logarithmic in the system
size. However, for any fixed $p>0$, the correction vanishes as $N\rightarrow\infty$, so that $\lim_{N\to\infty}\gamma_{\expval{r}}(p,N)=2$, recovering the universal RP localization threshold.
To characterize the finite-size dependence, we simulate a two-parameter ($(\grp,p)$) phase diagram in the $\expval{r}$ landscape. The color maps in FIG.~\ref{fig:rp_phase_diagram} show $\expval{r}$ for $N=1000$ and $N=4000$. To define a numerical crossover line for a fixed system size $N$, we use an anchored iso-$\expval{r}$ criterion. The reference value is chosen at the fully connected limit $p=1$ and at the finite-size crossover $\grp =\gamma_c=2$ as $r_*=\expval{r(\grp=2,p=1)}$. For each fixed value of $p$, we then determine the value of $\grp$ for which $\expval{r(\grp,p)}=r_*$. This numerical crossover line is shown by the black dashed curve in FIG.~\ref{fig:rp_phase_diagram}. The white curve shows the same by analytical estimation using Eq.~\eqref{eq:gamma_p_threshold}. The agreement between the two curves confirms that the localization boundary is shifted to smaller values of $\grp$ for a finite-size system when $p<1$ and the shift is logarithmic in system size. 
This happens because the displacement is controlled by the effective number of resonant hopping channels, $k_{\rm res}\sim pN^{1-\grp/2}$, which depends on both $p$ and $N$.

\section{Marked vertex search on RP graph}
\label{sec:search_RP_graph}

\subsection{Marked-vertex search via continuous-time quantum walk}
\label{sec:search_stnd_graph}

We first brief the standard continuous-time quantum-walk formulation of the marked-vertex or spatial-search problem~\cite{farhi1998quantum, ChildsPRA_2004}. A computational basis ${\ket{i}}_{i=1}^{N}$ spanning a Hilbert space of dimension $N$ can be used to denote the vertices of the graph $G$. The goal is to find a marked vertex $w$, represented by the state $\ket{w}$, starting from an initially delocalized state. In the standard spatial-search algorithm, the dynamics is governed by a Hamiltonian that encodes both the graph structure and the marked vertex. The search Hamiltonian is taken as
\begin{equation}
H_{\rm search} = \dyad{w} + \alpha H ,
\label{eq:stnd_search_hamil}
\end{equation}
where $H$ is a graph Hamiltonian defined as $H = \gamma_{\rm walk} L$ ($L$ is graph's Laplacian)~\cite{chakraborty_ctqw_qckt}. The parameter $\gamma_{\rm walk}$ denotes the transition probability rate between any two connected vertices~\cite{chakraborty_ctqw_qckt}. $\alpha$ controls the strength of the walk Hamiltonian relative to the oracle term, and the oracle term $\dyad{w}$ introduces a local perturbation that projects onto the marked vertex and attracts the dynamics towards $\ket{w}$. 

The initial state is chosen to be the uniform superposition over all vertices,
\begin{equation}
\ket{s}=\frac{1}{\sqrt{N}}\sum_{i=1}^{N}\ket{i}.
\label{eq:uniform_initial_state}
\end{equation}
This state contains no prior information about the position of the marked vertex and has initial marked-site probability $\abs{\braket{w}{s}}^2=\frac{1}{N}$. Under unitary evolution generated by $H_{\mathrm{search}}$, the state evolves as $\ket{\psi(t)} = e^{-i H_{\mathrm{search}} t} \ket{s}$ and the instantaneous success probability of finding the marked vertex is defined as
\begin{equation}
P_w(t) = \abs{\braket{w}{\psi(t)}}^2.
\label{eq:success_probability}
\end{equation}
The search is successful if the evolution amplifies $P_w(t)$ from its initial value $1/N$ to an order-one value at some later time. The peak success probability is defined as
\begin{equation}
P_{\rm peak} = \max_{0\leq t\leq T_{\max}} P_w(t),
\label{eq:peak_success_probability}
\end{equation}
where $T_{\max}$ is the maximum simulation time. The corresponding search time is defined as the earliest time at which this peak is reached,
\begin{equation}
t_{\rm peak}=\min\{t\in[0,T_{\max}]:P_w(t)=P_{\rm peak}\}.
\label{eq:peak_search_time}
\end{equation}
Thus, $P_{\rm peak}$ measures the reliability of finding the marked vertex upon measurement, while $t_{\rm peak}$ measures the time required to reach this maximal probability.

For graphs on which the standard spatial-search algorithm is optimal, one expects $P_{\rm opt}=\Theta(1)$ and $t_{\rm opt}=\Theta(\sqrt{N})$, giving a quadratic improvement over the classical $O(N)$ scaling~\cite{ChildsPRA_2004, sChakraborty_prl2016}. This behavior is known for several highly symmetric or sufficiently well-connected graphs, such as complete graphs, hypercubes, and high-dimensional lattices.

\subsection{An efficient algorithm}
\label{sec:CNR_algo}
The standard CTQW search algorithm described above requires a suitable choice of the hopping rate $\alpha$. For highly symmetric graphs, such as the complete graph or hypercube, this parameter can often be chosen analytically, and the search dynamics effectively reduces to a two-level oscillation between the initial state and the marked state. However, for a general graph there is no universal choice of $\alpha$~\cite{sChakraborty_prl2016, chakraborty_optimality_PRA}. The search performance depends sensitively on the spectral properties of the Hamiltonian that drives the walk, as well as on the overlap of its eigenstates with the marked vertex. This makes the direct application of the standard algorithm difficult for random general graphs. 

To address this issue, \citet{chakraborty_optimality_PRA} developed a spectral framework (we denote this by CNR after the authors) for characterizing the performance of spatial search by continuous-time quantum walks. In this approach, the graph Hamiltonian is normalized so that its eigenvalues lie in the interval $[0,1]$, and its spectral decomposition is written as
\begin{equation}
H = \sum_{i=1}^{N}\lambda_i \dyad{\lambda_i},
\qquad
\lambda_N=1>\lambda_{N-1}\geq \cdots \geq \lambda_1 .
\end{equation}
where $\ket{\lambda_i}$ are the eigenstates of the normalized graph Hamiltonian $H$ and $\lambda_i$ are the corresponding eigenvalues. The overlap between the marked vertex and the state corresponding to highest eigenvalue is denoted by $\epsilon = \abs{\braket{w}{\lambda_N}}^2$. For the standard Laplacian-based formulation on regular graphs, this highest-eigenstate description reduces to the usual uniform initial state $\ket{s}$, so that $\epsilon= \epsilon_{w_s}=\abs{\braket{w}{s}}^2=1/N$. 

The spectral gap between the two largest eigenvalues, $\Delta = \lambda_N-\lambda_{N-1} = 1-\lambda_{N-1}$, measures how well the principal eigenvalue $\lambda_N=1$ is separated from the rest of the spectrum~\cite{chakraborty_optimality_PRA}. However, this gap alone does not fully determine the performance of the continuous-time spatial-search algorithm~\cite{chakraborty_optimality_PRA}. The marked state $\ket{w}$ generally has nonzero overlap with several eigenstates of the walk Hamiltonian. The contribution of each eigenstate to the search dynamics depends both on its spectral separation from $\lambda_N=1$ and on its support at the marked vertex. Following the spectral framework of CNR, these contributions are collected through the overlap-weighted spectral quantities,
\begin{equation}
    S_k =\sum_{i\neq N}
    \frac{\abs{\braket{w}{\lambda_i}}^2}{(1-\lambda_i)^k},
    \qquad k \geq 1.
    \label{eq:Sk_CNR}
\end{equation}
Here, $k$ is a positive integer that specifies the power of the inverse spectral separation. Increasing $k$ gives progressively greater weight to eigenstates whose eigenvalues lie close to the principal eigenvalue. In particular, $S_1$ weights each eigenstate by the inverse of its spectral separation from $\lambda_N$, whereas $S_2$ weights the same separation quadratically and is therefore more sensitive to near-resonant eigenstates.

The numerator $\abs{\braket{w}{\lambda_i}}^2$ measures the support of the eigenstate $\ket{\lambda_i}$ on the marked vertex, while the denominator measures its spectral separation from the principal eigenvalue $\lambda_N=1$. An eigenstate therefore contributes most strongly when it has appreciable marked-state overlap and its eigenvalue lies close to $\lambda_N=1$. The quantities $S_1$ and $S_2$ therefore combine eigenvalue and eigenvector information from the full spectrum accessible to the marked state. Within the CNR framework, $S_1$ determines the critical hopping rate, while the ratio $S_1/\sqrt{S_2}$ and its inverse determine the achievable marked-state amplitude and the characteristic search time. These overlap-weighted spectral quantities therefore provide a more complete characterization of the search dynamics than the principal spectral gap ($\Delta$) alone.

In addition to the spectral sums suggested by ~\citet{chakraborty_optimality_PRA}, we also monitor the largest eigenstate weight on the marked vertex. This quantity is defined as
\begin{equation}
    \epsilon_{w_i} = \max_{i}
    \abs{\braket{w}{\lambda_i}}^2.
    \label{eq:max_overlap}
\end{equation}
The overlap $\epsilon_{w_i}$ measures how strongly the marked vertex is represented in the eigenbasis of the normalized Hamiltonian. A larger value of $\epsilon_{w_i}$ indicates that at least one eigenstate has substantial support on $\ket{w}$, which can enhance the probability accumulation at the marked vertex during the search dynamics.

Within this framework, the optimal hopping rate is determined by the first spectral sum,
\begin{equation}
\alpha=S_1.
\label{eq:CNR_critical_r}
\end{equation}
At this critical value, the dominant eigenstates of the search Hamiltonian (Eq.~\eqref{eq:stnd_search_hamil}) become nearly resonant and contain significant contributions from both the initial and marked-state components, enabling coherent amplitude transfer toward the marked vertex. The expected peak amplitude at the marked vertex and the corresponding search time are then controlled by $S_1$, $S_2$, and $\epsilon$. In particular, the maximum marked-state amplitude scales as $S_1/\sqrt{S_2}$. So the peak success probability is approximately
\begin{equation}
P_{\rm peak}
\sim
\frac{S_1^2}{S_2}.
\label{eq:CNR_peak_probability}
\end{equation}
The corresponding search time scales as
\begin{equation}
t_\ast
\sim
\frac{\pi}{2}
\frac{\sqrt{S_2}}{S_1\sqrt{\epsilon}} .
\label{eq:CNR_search_time}
\end{equation}
This spectral prescription is particularly useful for disordered graphs, where the search dynamics can vary strongly from one realization to another. Disorder modifies both the eigenvalue spacings and the spatial structure of the eigenstates. As a result, the success of marked-vertex search depends not only on how rapidly the walk spreads through the graph, but also on whether the eigenstates participating in the dynamics have appreciable overlap with marked vertex state $\ket{w}$. The CNR framework provides a way to relate the search observables directly to these spectral and eigenstate properties. In the following analysis, we therefore use the critical hopping rate determined from Eq.~\eqref{eq:CNR_critical_r} and examine how disorder affects the resulting peak success probability and search time.

\subsection{The disordered linegraph}
\label{sec:disord_linegraph}

To investigate the effect of disorder on CTQW-based marked vertex search, we first consider a simpler setting in which perhaps, the effect of disorder can be studied more directly. We take an open line graph with $N$ vertices, labelled by $i=0,1,\ldots,N-1$, and choose a marked vertex $w$. The line graph contains only nearest-neighbour edges. This provides a minimal geometry for examining how local disorder near the marked vertex modifies the spectral structure relevant for CTQW-based search.

We introduce disorder in two ways. First, the nearest-neighbour hopping amplitudes are allowed to fluctuate around a uniform background value $J_0$. The weighted adjacency matrix is defined through the Hadamard
product
\begin{equation}
    W = J \circ A,
    \label{eq:line_weighted_adjacency}
\end{equation}
where $J$ is the matrix of edge weights and $A$ is the adjacency matrix of the line graph. The edge weights are sampled according to
\begin{equation}
J_{mn} =
J_0+\eta_{mn},
\qquad
\eta_{mn}\sim \mathcal{N}\left(0,\; \sigma_{\rm line} N^{-\gamma_{\rm line}/2} \right) <J_0,
\label{eq:line_edge_weights}
\end{equation}
where $J_0$ is the uniform hopping, $\sigma_{\rm line}$ sets the strength of the hopping disorder and $\gamma_{\rm line}$ controls its scaling with system size. In our calculation, we have fixed $\sigma_{\rm line}=0.1$ and $\gamma_{\rm line} = 0.1$. 
\begin{figure}[t!]
  \centering
 \includegraphics[width=\linewidth]{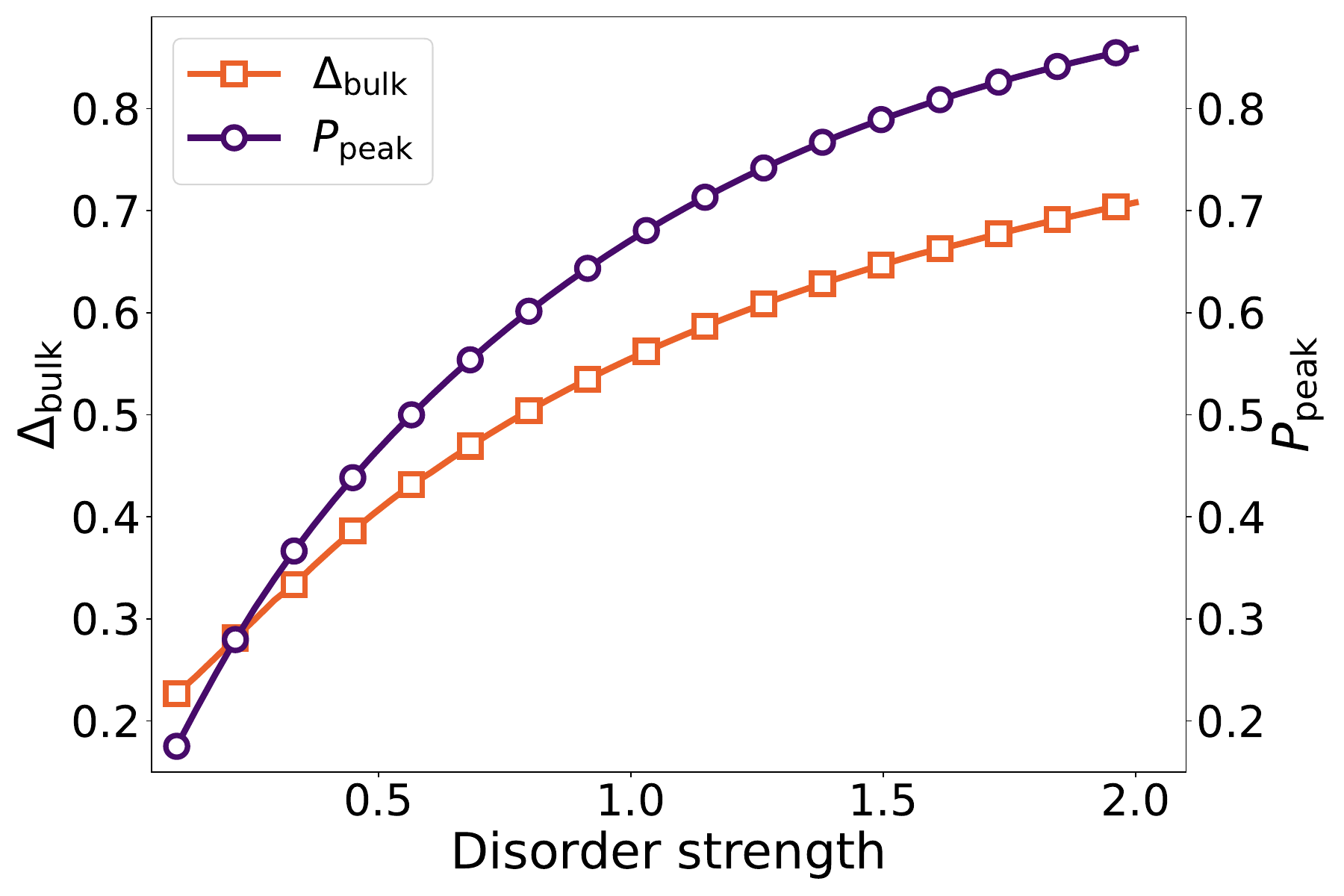}
 \caption{Variation of the peak search probability $P_{\mathrm{peak}}$ and spectral gaps $\Delta_{\mathrm{bulk}}=\lambda_{\max}-\lambda_{\mathrm{bulk}}$ with disorder strength for CTQW-based marked vertex search on a weighted open line graph.} 
  \label{fig:linegraph}
\end{figure}
Second, we add an onsite diagonal disorder locally around the marked vertex. Let $R_{\rm dis}$ denote the range of disorders. The onsite potential $V_m$ is taken to be nonzero only for vertices within distance $R_{\rm dis}$ of the marked vertex with
\begin{figure}[t!]
    \centering
    \includegraphics[width=0.9\linewidth]{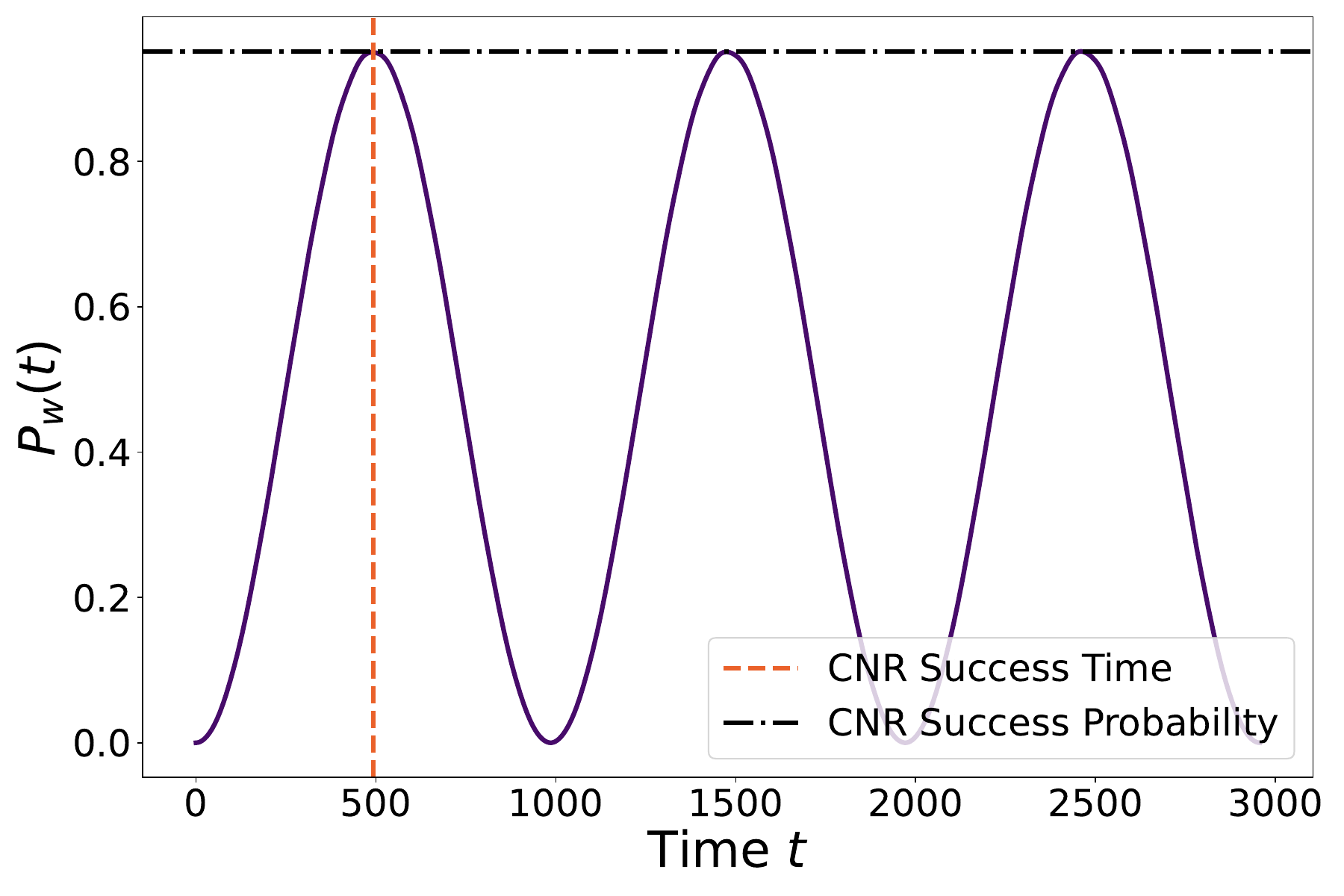}
    \caption{Time evolution of the marked-vertex success probability $P_w(t)$. The solid curve shows the numerical CTQW dynamics. The red dashed line marks the CNR-predicted first-peak search time, while the black dash-dotted line denotes the corresponding CNR-predicted peak success probability.}
    \label{fig:prob_osc}
\end{figure}
\begin{equation}
V_m\sim \mathcal{N}(0,\sigma_{\rm loc}^2).
\label{eq:local_onsite_disorder}
\end{equation}
Here $\sigma_{\rm loc}$ is the strength of the local disorder. The marked vertex itself is not included in the onsite disorder region. This choice keeps the oracle term distinct from the background disorder and allows us to study how disorder in the neighbourhood of the marked site affects the search dynamics.

The weighted degree matrix is defined by $D_{mn}=\delta_{mn}\sum_{\ell=0}^{N-1} W_{m\ell}$. The disordered weighted Laplacian is then $\widetilde{L} = D-W+V$, where $V_{mn}=V_m\delta_{mn}$. For CTQW dynamics, we use
\begin{equation}
H_{\rm line}=-\widetilde{L}.
\label{eq:H_line}
\end{equation}

We then apply the spectral search prescription (on search Hamiltonian $H_{\rm search,\:line} = \dyad{w} + rH_{\rm line}$ similar to Eq.~\eqref{eq:stnd_search_hamil}) following CNR algorithm as discussed in detail in section~\ref{sec:CNR_algo}. To quantify how local disorder modifies the spectral structure controlling the search, we monitor the separation of the state corresponding to the highest eigenvalue from the rest of the spectrum. We define a bulk spectral separation as $\Delta_{\rm bulk}=\lambda_{\max}-\lambda_{\rm bulk},$ where $\lambda_{\rm bulk}$ denotes the largest eigenvalue belonging to the dense spectral band below $\lambda_{\max}$. The search performance is quantified by the peak probability (see Eq.~\eqref{eq:peak_success_probability}).

FIG~\ref{fig:linegraph} shows the dependence of the spectral gaps and the peak search probability on the local disorder strength $\sigma_{\rm loc}$. We find that increasing the local disorder enhances the separation $\Delta_{\rm bulk}$ and simultaneously increases the peak probability of detecting the marked vertex. This suggests that a local disorder can create a more isolated spectral structure near the marked site, thereby increasing the overlap between $\ket{w}$ and the eigenstates that dominate the search dynamics. As the relevant state becomes more separated from the bulk the coherent population transfer to the marked vertex becomes more efficient, leading to a larger success probability. 

This simple line graph analysis shows that disorder need not always suppress the search for the marked vertex. Instead, local disorder can enhance the detectability of the marked vertex by increasing spectral separation and concentrating eigenstate weight near the marked region. However, the line graph is a highly constrained one-dimensional geometry, and the disorder introduced here is local rather than fully network-wide. Therefore, while this example provides useful intuition, it is not sufficient to characterize search in general disordered graphs. This motivates the more extensive analysis of graph-constrained disordered Hamiltonians in the following subsection.

\subsection{Marked-vertex search in the RP--ER ensemble}
\label{sec:search_tradeoff}

\begin{figure}[t!]
\centering
\includegraphics[width=0.9\linewidth]{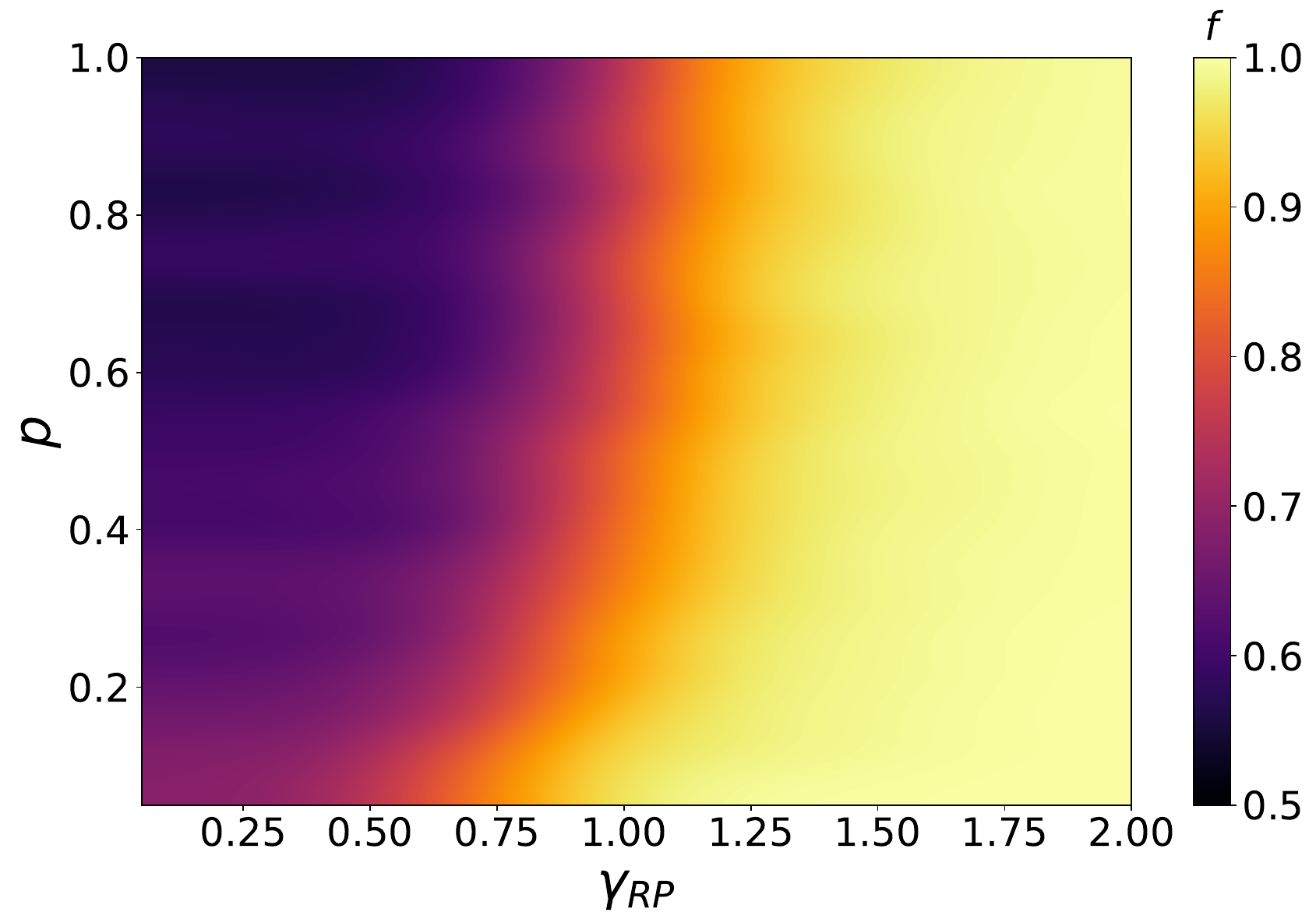}
\caption{Fraction of successful realizations in the $(\gamma_{\mathrm{RP}},p)$ plane, where a realization is counted as successful if the first-peak probability satisfies $P_{\mathrm{peak}}\geq 0.8$. Here $N=1000$.}
\label{fig:search_phase}
\end{figure}

We now investigate how RP-type disorder affects marked-vertex search on the graph-constrained RP ensemble as discussed in Section~\ref{sec:RP_on_graphs}. For each realization, the system is initialized in the principal eigenstate of the normalized hopping Hamiltonian $H_0$, and the marked vertex is introduced through the search Hamiltonian (see Eq.~\eqref{eq:stnd_search_hamil})  where $\alpha$ is tuned according to the spectral prescription discussed in section~\ref{sec:CNR_algo}. For each realization, the underlying graph topology (\textit{i.e.,} ER background) is kept fixed while only the RP disorder is varied randomly. This allows us to isolate the effect of RP-type Hamiltonian disorder on the marked-vertex search dynamics. We have also repeated the analysis in a doubly random setting, where both the graph realization and the RP disorder are varied, and obtained qualitatively similar behavior; these results are presented in Appendix~\ref{app:double_random}.

For every realization, we record the first maximum of the probability of success $P_w(t)$, which is defined in Eq.~\eqref{eq:success_probability}. The corresponding time is denoted by $\tau$, and the first-peak success probability is denoted by $P_{\mathrm{peak}} = P_w(\tau)$.

As illustrated in FIG.~\ref{fig:prob_osc}, the predicted CNR search time and success probability (see Eq.~\eqref{eq:CNR_peak_probability} and \eqref{eq:CNR_search_time}) agree well with the first oscillation peak of our numerical dynamics. We classify a realization as successful when $P_{\mathrm{peak}} \geq 0.8 $, and define the success fraction $f$ as the fraction of disorder realizations that meet this condition. 

FIG.~\ref{fig:search_phase} shows the resulting success fraction in the $(\grp,p)$ plane. For small $\grp$, the system is in the ergodic regime and the success fraction is comparatively low. In this regime, the marked state has only a weak overlap with individual eigenstates, so the probability amplitude remains distributed across many modes and does not concentrate efficiently on the marked vertex. As $\grp$ increases, $f$ increases and eventually approaches unity over a wide range of $p$. Thus, stronger RP disorder improves the probability of detecting the marked vertex.

\begin{figure}[t!]
\centering
\includegraphics[width=\linewidth]{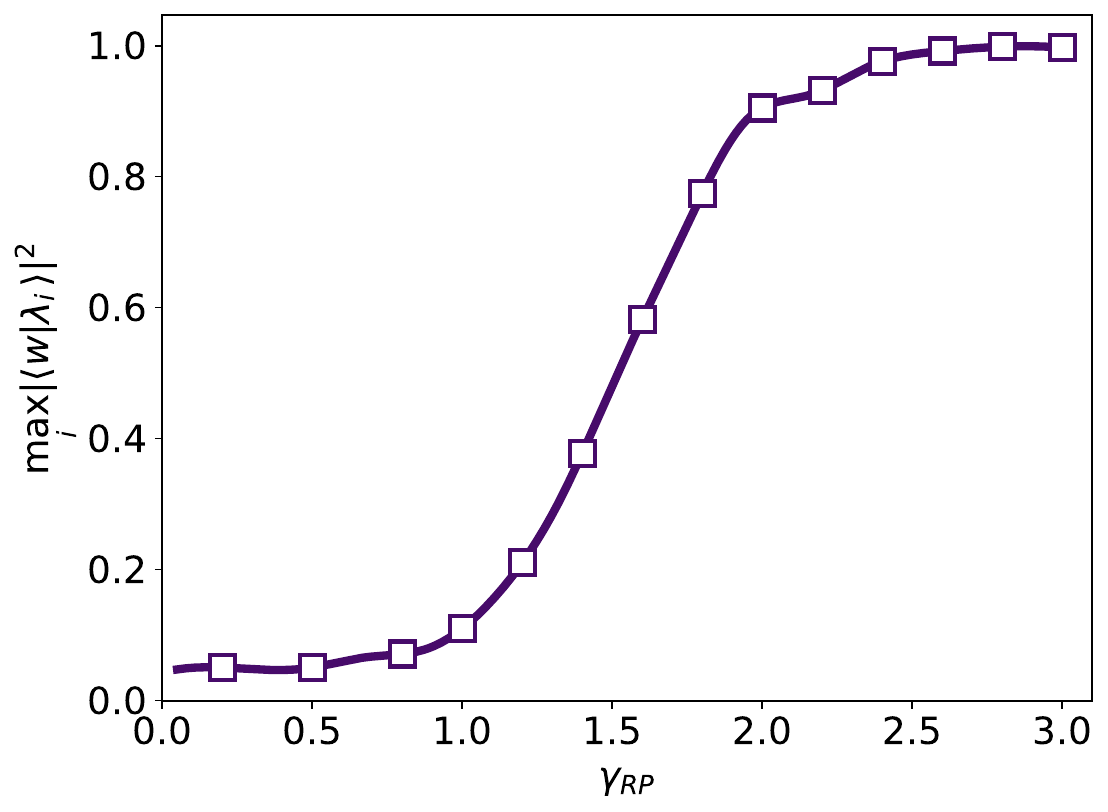}
\caption{Maximum marked-state overlap $\epsilon_{w_i}=\max_i|\braket{w}{\lambda_i}|^2$ as a function of $\gamma_{\mathrm{RP}}$ for $N=1000$. The rapid increase of $\epsilon_{w_i}$ signals the localization of the marked vertex onto a small number of eigenstates.}
\label{fig:max_overlap}
\end{figure}

The origin of this improvement can be seen from the maximum marked-state overlap $\epsilon_{w_i}$ (see Eq.~\eqref{eq:max_overlap}) shown in FIG.~\ref{fig:max_overlap}. In the ergodic regime, $\epsilon_{w_i}$ remains small, indicating that the marked vertex is distributed over many delocalized eigenstates. With increasing $\grp$, $\epsilon_{w_i}$ grows rapidly and approaches unity, showing that the marked vertex becomes strongly associated with a small number of localized eigenstates. This localization enhances the ability of the search dynamics to build up probability on $\ket{w}$, which explains the increase of the first-peak success probability.

\begin{figure}[t!]
\centering
\includegraphics[width=\linewidth]{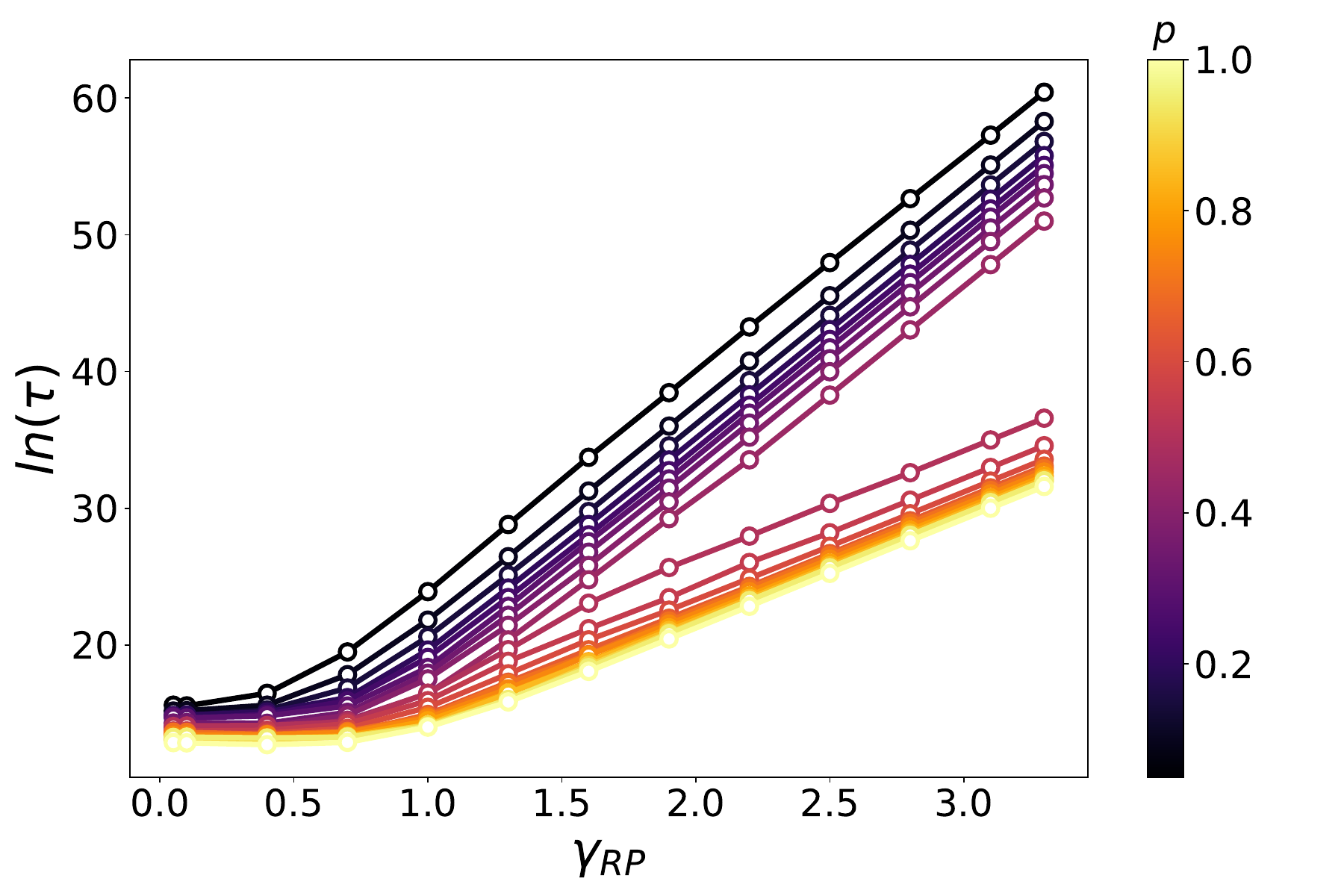}
\caption{Median first-peak search time $\ln(\tau)$ as a function of $\grp$ for different $p$ at $N=1000$. The search time increases with $\grp$, indicating progressively slower dynamics as the system moves toward the localized regime. For fixed $\grp$, denser graphs, \textit{i.e.} graphs with larger $p$, generally exhibit shorter search times.}
\label{fig:search_time}
\end{figure}

However, this improvement comes at the cost of a longer runtime. FIG.~\ref{fig:search_time} shows that the median search time $\ln(\tau)$ increases strongly with $\grp$. This behavior reflects the fact that the search dynamics is not controlled only by the largest spectral gap---instead, in the CNR framework, it is governed by the overlap-weighted spectral sums (see Eq.~\eqref{eq:Sk_CNR}) where $\lambda_N=1$ is the upper spectral edge. These quantities combine two pieces of information, how close each eigenvalue lies to the upper spectral edge and how strongly the corresponding eigenvector overlaps with the marked vertex.

FIG.~\ref{fig:max_overlap} clarifies why the runtime increases in the localized regime. Although increasing $\grp$ produces a large overlap with the marked vertex, this overlap is carried mainly by localized eigenstates that are not necessarily close to the upper spectral edge $\lambda=1$. Therefore, the weight of marked-state becomes concentrated, but the effective spectral distance relevant for the search dynamics also increases. Since the characteristic search time scales as $\sqrt{S_2}/S_1$ (see Eq.~\eqref{eq:CNR_search_time})---up to the overall factor set by the initial marked-state overlap---a redistribution of marked-state weight away from the spectral edge naturally leads to slower oscillations and hence larger $\tau$.

In a nutshell, these results demonstrate a clear trade-off between success probability and runtime. In the weak-disorder ergodic regime, the dynamics is faster, but the marked-state probability span on many eigenstates, resulting in a lower fraction of success. In the localized regime with strong-disorder, the marked vertex has a much larger overlap with localized eigenstates, leading to high first-peak success probability; however, the relevant spectral sums produce a longer characteristic search time. We find these results to be consistent across the system sizes ($N=1000$, $2000$, and $4000$) considered. Thus, RP disorder improves the reliability of marked-vertex detection, but only at the cost of slowing down the CTQW search dynamics.

\section{Conclusion}
\label{sec:conclusion}
In this work, we studied continuous-time quantum-walk-based marked-vertex search on disordered random graphs. Our main objective was to understand how controlled disorder on random graphs affect the ability of a quantum walker to find a marked vertex. We introduced a graph-constrained Rosenzweig--Porter (RP) ensemble on Erd\H{o}s--R'enyi (ER) random graphs. In this construction, the ER edge probability $p$ controls the availability of hopping channels, while the RP parameter $\grp$ controls the strength of the allowed random couplings. This provides doubly random framework in which network sparsity and disorder strength can be tuned independently.

We introduce a graph-constrained version of RP and first established that the characteristic RP spectral crossover survives under graph constraints in section~\ref{sec:RP_on_graphs}. By analyzing the disorder-averaged adjacent-gap ratio $\expval{r}$, we found a clear crossover from Wigner--Dyson-like statistics at small $\grp$ to Poisson-like statistics at large $\grp$. This behavior persists from sparse ER graphs to the fully connected limit. We further derived a simple analytical estimate for the finite-size localization boundary based on a resonant-hybridization argument. The resulting expression Eq.~\eqref{eq:gamma_p_threshold}, captures the observed shift of the apparent crossover with graph sparsity. It also recovers the conventional RP localization threshold $\grp=2$ in the fully connected limit $p=1$, and approaches the same threshold in the thermodynamic limit for any fixed $p>0$. Thus, graph sparsity mainly produces a finite-size suppression of hybridization, while the asymptotic localization threshold remains consistent with RP physics.

Before moving to the full graph-constrained RP ensemble, we first examined a disordered line graph for understanding how disorder modifies CTQW-based marked-vertex search in section~\ref{sec:disord_linegraph}. In this controlled case, we observed that onsite disorder can enhance the peak probability by increasing the marked-state overlap, while also modifying the relevant spectral gaps and modifying the search dynamics. The disordered line graph therefore provided a useful platform showing that search performance is controlled not only by transport, but also by the spectral weight carried by eigenstates overlapping with the marked vertex. However, a line graph is a very trivial setting, which motivated us to move to the more general RP--ER setting, where we have independent control over graph sparsity and Hamiltonian disorder.

We then used the graph-constrained RP ensemble to study marked-vertex search. Using the spectral search CNR framework~\cite{chakraborty_optimality_PRA} (as detailed in section~\ref{sec:CNR_algo}), we tuned the hopping rate through the overlap-weighted spectral sum $S_1$ and related the search performance to the spectral quantities $S_1$, $S_2$, and the marked-state overlap structure. Our results in section~\ref{sec:search_tradeoff} show a clear trade-off between success probability and runtime. In the weak-disorder, ergodic regime, the quantum walk spreads rapidly through the graph and the first-peak search time is relatively short. However, the marked-state weight is distributed over many eigenstates, leading to lower peak success probability. In contrast, as $\grp$ is increased and the system becomes localized, the marked vertex develops a large overlap with a small number of eigenstates. This strongly enhances the peak success probability and increases the fraction of successful realizations. The cost of this enhancement is a significantly longer search time, because the relevant overlap weight is no longer carried primarily by eigenstates close to the upper spectral edge. Therefore, localization improves the reliability of detecting the marked vertex, but slows down the coherent search dynamics.

Our results should be contrasted with recent studies of localization and multifractality on random graphs. \citet{SahaRoy_PRB2026} studied Anderson localization on spatially structured random graphs, where hopping amplitudes decay exponentially with graph distance. Their focus was the localization phase diagram generated by the competition between hopping range and onsite disorder, and they found a direct transition between delocalized and localized phases without evidence of an intervening multifractal phase. \citet{Cugliandolo_PRB2024} studied weighted adjacency matrices of sparse ER graphs and identified a weak multifractal phase arising from graph-topology heterogeneity and fluctuations in the hopping amplitudes. In comparison, our model is not based on distance-dependent hopping or on a fixed finite average degree weighted ER ensemble. Instead, we impose RP scaling on ER-constrained couplings, so that the disorder strength is controlled by $\grp$ while the graph sparsity is independently controlled by $p$. More importantly, our central observable is not only the localization phase diagram, but the performance of a quantum search algorithm. \citet{Cattaneo_PRA2018} studied CTQW spatial search under dynamical link noise and showed that graph topology and target-node connectivity strongly affect robustness against noise. However, their work does not address static Hamiltonian disorder or localization-driven changes in the search spectrum. We show that the same spectral and eigenvector structures that characterize the RP crossover also directly control the algorithmic quantities $P_{\mathrm{peak}}$, $\tau$, and the success fraction $f$.

There are several natural directions for future work. First, it would be useful to study the scaling of the optimal search time and success probability with system size more systematically across the ergodic, nonergodic extended, and localized regimes. Nonergodic extended regime may provide an intermediate balance between fast transport and enhanced marked-state overlap. Second, the present single-marked-vertex setting can be extended to multiple marked vertices, where the interplay between disorder, graph connectivity, and the spatial arrangement of the marked set may lead to qualitatively different search behavior.

\section{Acknowledgment}
SC acknowledge the support of the Prime Minister’s Research Fellowship (PMRF). The authors would also like to acknowledge \emph{Paramshakti} Supercomputer facility at IIT Kharagpur---a national supercomputing mission of the Government of India, for providing the necessary high-performance computational resources. T\v{C} was supported by an appointment to the JRG Program at the APCTP through the Science and Technology Promotion Fund and Lottery Fund of the Korean Government and by the Korean Local Governments - Gyeongsangbuk-do Province and Pohang City. RKR acknowledges the support by the Institute for Basic Science in Korea (IBS-R024-D1) and particularly the hospitality of PCS--IBS. Much of the idea behind this work was conceived during the visit.

\appendix
\section{Doubly random graph and disorder realizations}
\label{app:double_random}

\begin{figure}[ht]
\centering
\includegraphics[width=\linewidth]{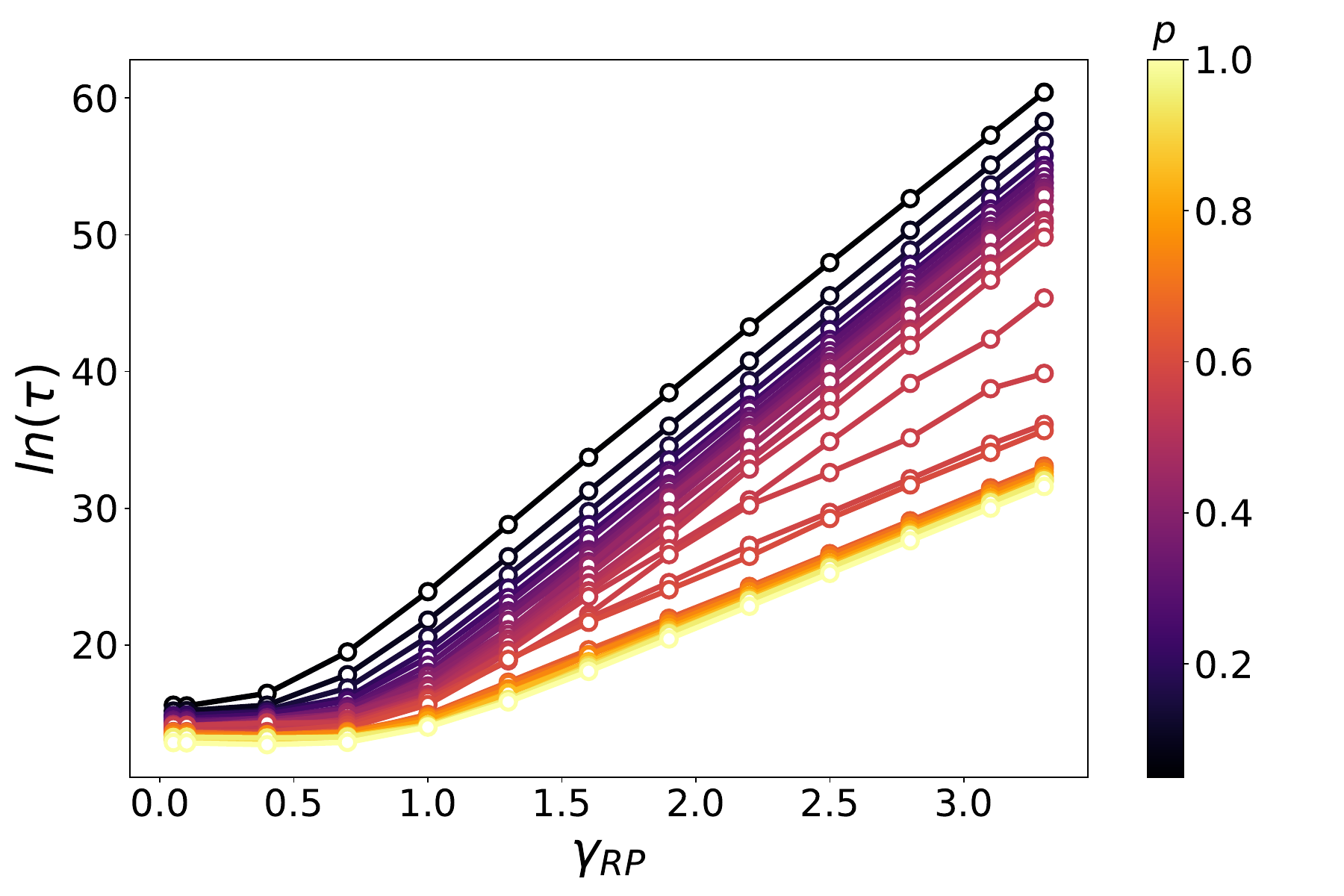}
\caption{Median first-peak search time $\ln(\tau)$ as a function of $\grp$ for different edge probabilities $p$ at $N=1000$ in the doubly random ensemble. For each sample, both the ER graph realization and the RP disorder realization are generated independently, and the marked vertex is chosen randomly. Similar to the Fig.~\ref{fig:search_time}, the search time increases with $\grp$, while denser graphs generally exhibit shorter search times.}
\label{fig:RP_ER_doublyrndm}
\end{figure}

In the Section~\ref{sec:search_tradeoff}, the marked-vertex search results are obtained by fixing the underlying ER graph for each value of $(N,p)$ and varying only the RP disorder. This choice isolates the effect of RP-type Hamiltonian disorder on the search dynamics. As a robustness check, we also consider a doubly random setting in which both the ER graph realization and the RP disorder are independently resampled for each realization. The marked vertex is also chosen randomly in each realization.

For every pair $(\grp,p)$, we construct an independent graph-constrained RP Hamiltonian, normalize the hopping Hamiltonian, initialize the system in its principal eigenstate, and tune the hopping rate using the spectral prescription described in Sec.~\ref{sec:CNR_algo}. We then compute the first-peak search time $\tau$ and report the median value of $\ln(\tau)$ over the ensemble.

Figure~\ref{fig:RP_ER_doublyrndm} shows that the qualitative behavior remains the same as in the fixed-graph analysis. The search time increases systematically with increasing $\grp$, indicating slower dynamics as the system approaches the localized regime. The dependence on graph connectivity is also preserved; denser graphs generally support faster search, while sparse graphs lead to longer runtimes. Thus, the trade-off observed in the main text between enhanced marked-state localization and slower search dynamics is not an artifact of fixing the graph topology. Instead, it persists when both sources of randomness, graph connectivity and RP disorder, are varied simultaneously.

\bibliography{references.bib}

\end{document}